\documentclass[12pt]{article}

\usepackage{amssymb}
\usepackage[american]{babel}

\def\dfrac#1#2{\displaystyle{\frac{#1}{#2}}}

\def\frak{{\bf }}
\def\mathbb{{\bf }}

\let\d=\delta
\let\e=\varepsilon
\let\l=\lambda

\let\m=\mu
\let\n=\nu
\let\P=\Pi

\let\x=\xi

\let\y=\psi
\let\z=\zeta

\def\nth{n^{\underline{{\rm th}}}}

\let\ot=\otimes

\def\cC{{\cal C}}
\def\cM{{\cal M}}
\def\cS{{\cal S}}
\def\cV{{\cal V}}

\def\II{{\mathbb I}}
\def\RR{{\mathbb R}}

\def\bbR{{\mathbb R}}

\def\NN{{\mathbb N}}

\def\tr{{\rm \,tr\,}}
\def\Tr{{\rm \,Tr\,}}

\def\det{{\rm \,det\,}}

\def\End{{\rm \,End\,}}

\def\vol{{\rm \,vol\,}}

\def\const{{\rm \,const\,}}
\def\be{\begin{equation}}
\def\ee{\end{equation}}
\def\bea{\begin{eqnarray}}
\def\eea{\end{eqnarray}}
\def\bes{\begin{displaymath}}
\def\ees{\end{displaymath}}

\newtheorem{theorem}{Theorem}

\newtheorem{remark}{Remark}

\def\sideremark#1{\ifvmode\leavevmode\fi\vadjust{\vbox to0pt{\vss
 \hbox to 0pt{\hskip\hsize\hskip1em
 \vbox{\hsize2cm\tiny\raggedright\pretolerance10000
 \noindent #1\hfill}\hss}\vbox to8pt{\vfil}\vss}}}%
\newcommand{\nnn}[1]{(\ref{#1})}

\let\a=\alpha
\let\b=\beta
\let\f=\varphi
\let\g=\gamma
\let\N=\nabla

\def\cR{{\cal R}}

\def\ci{C^\infty}
\let\N=\nabla

\def\auto{{\mbox{Aut}}}
\def\endo{{\mbox{End}}}
\def\id{{\mbox{id}}}
\def\LC{{\rm LC}}
\def\orth{{\rm O}}
\def\pin{{\rm Pin}}
\def\so{{\rm SO}}
\def\spin{{\rm Spin}}

\begin{document}

\begin{titlepage}

\null
\vspace{2cm}
\centerline{\LARGE\bf Heat Kernel Asymptotics of Operators}
\medskip
\centerline{\LARGE\bf with Non-Laplace Principal Part}
\bigskip
\bigskip
\centerline{\Large\bf Ivan G. Avramidi}
\bigskip
\centerline{\it Department of Mathematics, The University of Iowa}
\centerline{\it Iowa City, IA 52242, USA}
\medskip
\centerline{\rm and}
\medskip
\centerline{\it Department of Mathematics}
\centerline{\it New Mexico Institute of Mining and Technology} 
\centerline{\it Socoro, NM 87801, USA} 
\bigskip
\centerline{\rm and}
\medskip
\centerline{\Large\bf Thomas Branson}
\bigskip
\centerline{\it Department of Mathematics, The University of Iowa}
\centerline{\it Iowa City, IA 52242, USA}
\bigskip
\bigskip
\centerline{October 8,2000}
\medskip
\vfill

\begin{quote}
We consider second-order elliptic partial differential operators
acting on sections of vector bundles over a compact Riemannian manifold without
boundary, working {\em without} the assumption of Laplace-like principal part
$-\N^\mu\N_\mu\,$.  Our objective is to obtain information on the asymptotic
expansions of the corresponding resolvent and the heat kernel.  The heat kernel
and the Green's function are constructed explicitly in the leading order.  
The first two coefficients of the heat kernel asymptotic expansion are computed
explicitly. A new semi-classical ansatz as well as the complete recursion
system for the heat kernel of non-Laplace type operators is constructed.
Some particular cases are studied in more detail.  
\end{quote}

\end{titlepage}

\baselineskip=18pt
\section{Introduction}
\setcounter{equation}0

The resolvent and the heat kernel of elliptic partial differential operators are
of great importance in mathematical physics, differential geometry and quantum
theory \cite{hadamard23,gilkey95,berline92,dewitt84,maslov76}.  Of special
interest in the study of elliptic operators are the so-called heat kernel
asymptotics.  It is well known \cite{gilkey95} that for a second-order,
elliptic,  self-adjoint partial differential operator $F$ with a positive
definite leading symbol, acting on sections of a vector bundle over a compact,
boundariless manifold $M$ of dimension $m$, an asymptotic expansion of the
following form is valid as $t\downarrow 0$:
\bes
\Tr_{L^2}\exp(-tF)
\sim (4\pi t)^{-m/2}\sum_{k\ge 0}{(-t)^k\over k!} A_k.
\ees
The coefficients $A_k$ are called the {\em heat invariants}, or 
{\em heat kernel
coefficients}.  They are spectral invariants of the operator $F$, and encode
information about the asymptotic properties of the spectrum. 
Note that our normalization of the coefficients $A_k$ differs by the factor
$(-1)^k/k!$ from that used in \cite{gilkey95}. This normalization has been used
in previous works of one of the authors (see the book \cite{avra00}, the papers
\cite{avra91b,avra99,avra98} and others). It has the
advantage that for Laplace type operators with a potential (see definition
below), $F=-\Delta+q$, the numerical coefficient of the term $\int_M d\vol(x)
\tr\, q^k$ in $A_k$ is equal to $1$ for any $k$.

An important subclass of the class of operators described above are the
operators of {\em Laplace type}:  those for which the leading symbol takes the
form $g^{\mu\nu}\xi_\mu\xi_\nu\,$, where $g^{\mu\nu}$ is a non-degenerate,
positive definite metric on the cotangent bundle of $M$.  For such operators,
the leading symbol naturally defines a Riemannian metric on $M$ (the inverse
$g_{\mu\nu}$ of the leading symbol).  Alternatively, one may take the 
Riemannian
metric as given, and produce the 
many natural operators of Laplace type which are so
important in Physics.

The assumption of Laplace type affords a considerable simplification in the
study of spectral asymptotics.  Partly as a result of this, much is known about
the resolvent, the heat kernel, and the zeta function in this category of
operators \cite{gilkey95,berline92,avra91b,avra99,avra00}.  In particular, the
heat kernel coefficients $A_k$ for Laplace type operators are known {\it
explicitly} up to $k=4$ \cite{avra91b,avra00} (for a review, see
\cite{avra99}).

In this paper, we take a Riemannian metric as given, and study the most general
class of second-order operators $F$, acting on sections of a vector bundle
$\cV$, with positive definite leading symbol.  That is, we drop the assumption
of Laplace type, and assume only that
\bes
\sigma_2(F;x,\xi)=a^{\mu\nu}(x)\xi_\mu\xi_\nu\,,
\ees
where $a^{\mu\nu}$ is a symmetric two-tensor valued in $\endo(\cV)$.  (We do
{\em not} assume that $a^{\mu\nu}=g^{\mu\nu}\id_{\cV}\,$, nor that $a^{\mu\nu}$
is factored as $g^{\mu\nu}E$ for $E$ a section of $\auto(\cV)$.)  We shall
sometimes call these ``NLT'' (for ``non-Laplace type'') operators.  Of course,
Laplace type is a special case, so to be more precise, NLT operators are
operators that are not {\em necessarily} of Laplace type.
NLT operators arise naturally in such areas of mathematical
physics as quantum gauge field theory and quantum gravity
\cite{dewitt84,avra00}, differential geometry, classical continuum mechanics
\cite{kupradze65} and others.

The most elementary examples that one can use to illustrate the class of
operators in question are the {\em weighted form Laplacians}.  If $d$ is the
exterior derivative and $\delta$ its formal adjoint, the {\em form Laplacian} 
is
$\Delta=\delta d+d\delta$; this differs from the {\em Bochner Laplacian}
$g^{\mu\nu}\N_\mu\N_\nu$ by an operator of order zero, the so-called {\em
Bochner-Weitzenb\"ock} operator.  If $a,b$ are real constants, an operator of
the form $a\delta d+bd\delta$ may be termed a weighted form Laplacian; such
operators are elliptic if $a\ne 0\ne b$, and have positive definite leading
symbol if $a,b>0$.

%
The non-Laplace type operators on differential forms  have been studied
extensively by many authors under the general name ``nonminimal operators'',
or ``exotic operators'' (see
\cite{gilkey91,gilk-bransMN,gusynin91,gusynin95,gusynin99,cho95,alex96} and
references therein). In particular, the paper \cite{gilkey91} contains a
complete discussion of the coefficients $A_k$ of the weighted form Laplacian
for all $k$ and of $A_0$ and $A_1$ of the operator with a potential function
(Th. 1.2 and Th 1.3, pp. 2089-2090). In papers
\cite{gusynin91,gusynin95,gusynin99} a computer algorithm has been developed
that employs the calculus of pseudodifferential operators and the
coefficients $A_0$, $A_1$ and $A_2$ (more precisely, the local
(non-itegrated) coefficients $a_0$, $a_1$ and $a_2$) have been computed.

The $\zeta$-function for non-Laplace type operators acting on symmetric
$2$-tensors have been studied in \cite{cho95} (for the origin of such
operators, see \cite{dewitt84,avra00}). By restricting to maximally symmetric
even dimensional manifolds the authors were able to compute the eigenvalues of
such operators and  evaluate the $\zeta$-function at zero, $\zeta(0)$, (which
is, in fact, equivalent to the coefficient $A_{m/2}$ \cite{gilkey95}) for a
certain special case of the operator in Euclidean spaces and $m$-spheres  in 
dimensions  $m=2,4,6,8,10$.

In general, the study of  spin-tensor
quantum gauge fields in arbitrary gauge necessarily leads to non-Laplace type
operators acting on sections of {\it general tensor-spinor bundles}. It is
precisely these operators that are of prime interest in the present paper. The
main examples here are the symmetric 2-tensor bundle (gravitational field,
spin 2) and the spin-vector bundle (gravitino field, spin 3/2) (for a
discussion of such operators in gauge field theories, see \cite{avra99b}).

\medskip
Let us formulate our main result from the very beginning.

\begin{theorem} Let $F:\ C^\infty({\cal V})\to C^\infty({\cal V})$ be a
self-adjoint elliptic second-order partial differential operator with a 
positive
definite leading symbol, acting on sections of a tensor-spinor bundle 
${\cal V}$ of
fiber dimension $d$ over a compact manifold $M$ of dimension $m$ without
boundary.  Let $F=-a^{\mu\nu}\nabla_\mu\nabla_\nu+q$, where
$\nabla$ is a connection on the vector bundle ${\cal V}$, $a^{\mu\nu}$ is a
parallel symmetric two-tensor valued in $\End({\cal V})$ and $q$ is an
endomorphism of the bundle ${\cal V}$.  The curvature of the connection 
$\nabla$
is defined by $[\nabla_\mu, \nabla_\nu] \varphi = R^\alpha{}_\beta{}_{\mu\nu}
T^\beta{}_\alpha \varphi$, where $R^\alpha{}_{\beta\mu\nu}$ is the Riemann
curvature tensor and $T^\alpha{}_\beta$ is given by
the representation of ${\frak{so}}(m)$ which induces the bundle $\cV$.
Let $\xi\in T^*M$ be a
cotangent vector, and let 
$\lambda_i(\xi)=\mu_i|\xi|^2$, $(i=1,\cdots,s)$, $\mu_i>0$, be
the eigenvalues of the leading symbol, $\sigma_2(F) = a^{\mu\nu}
\xi_\mu\xi_\nu$, with the multiplicities $d_i$.
Then the corresponding (orthogonal) eigenspace projections have the form
\bea
\Pi_i(\xi) &=& \sum_{n=0}^p
\Pi_{i(2n)}^{\mu_1\cdots\mu_{2n}} {1\over |\xi|^{2n}}
\xi_{\mu_1} \cdots \xi_{\mu_{2n}}
\nonumber\\
&=&\sum_{k=1}^s c_{ik}a^{(\mu_1\mu_2}\cdots a^{\mu_{2k-3}\mu_{2k-2})}
{1\over |\xi|^{2k-2}}\xi_{\mu_1} \cdots \xi_{\mu_{2k-2}}\;,
\nonumber
\eea
where the numbers
$s$ and $p$ depend on the structure of the bundle ${\cal V}$ and the leading
symbol.
The matrix of coefficients $\cC:=(c_{ik})$
is inverse to the (Vandermonde) matrix of powers $\cM=(\kappa_{kj})
:=(\m_j^{k-1})$.

Furthermore, the $L^2$ trace of the heat kernel has the following asymptotics
as $t\to 0$
$$
\Tr_{L^2}\exp(-tF)=(4\pi t)^{-m/2}\left[A_0-tA_1+O(t^2)\right].
$$
The coefficients $A_0$ and $A_1$ are defined by
$$
A_0=\sum_{i=1}^s d_i\mu_i^{-m/2}\vol(M)\,,
$$
$$
A_1=\int_M d\vol(x) \left\{\tr_{\cal V}\left(a_0q\right)+\beta\,R\right\}\,,
$$
where $R$ is the scalar curvature,
$$
a_0=\sum_{i=1}^s \mu_i^{-m/2}<\Pi_i>\,,
$$
\bea
<\Pi_i>=\Pi_{i(0)}&=&
\sum_{k=1}^s c_{ik}{\Gamma(m/2) (2k-2)!\over \Gamma(m/2+k-1) 2^{2k-2}(k-1)!}
\nonumber\\
&&\times g_{(\mu_1\mu_2}\cdots g_{\mu_{2k-3}\mu_{2k-2})}
a^{(\mu_1\mu_2}\cdots a^{\mu_{2k-3}\mu_{2k-2})}\,,
\eea
and $\beta$ is a constant defined by
\bea
\beta&=&
-{1\over 6m}\sum_{i=1}^s \mu^{-m/2}_i\tr_{\cal V}
\left(<\Pi_i>g_{\mu\nu}a^{\mu\nu}\right)
\nonumber\\[11pt]
&&
+{1\over 12(m-1)}
\sum_{1\le i,k\le s;\,i\ne k}
\left[\kappa_{ik}{(3m-2)}
+4(m+2)\mu_i\sigma_{ik}\right]
\tr_{\cal V}\left<\Pi_i\tilde J^\alpha \Pi_k\tilde J_\alpha\right>
\nonumber\\[11pt]
&&
-{1\over 2(m-1)}\sum_{1\le i,k\le s;\,i\ne k}
\kappa_{ik}\tr_{\cal V}
\left\{\left[\left<\tilde J^\nu\Pi_i \tilde J^\alpha \Pi_k\right>
+\left<\Pi_i \tilde J^\alpha \Pi_k \tilde J^\nu\right>\right]
T_{[\nu\alpha]}\right\}\,,
\nonumber
\eea
with
$$
\kappa_{ik}=-{\Gamma(m/2-1)\over 2\Gamma(m/2+1)}
\left\{m\,{\mu_i^{-m/2}+\mu_k^{-m/2}\over (\mu_i-\mu_k)}
+2\,{\mu_i^{-m/2+1}-\mu_k^{-m/2+1}\over (\mu_i-\mu_k)^2}
\right\},
\label{9144xxx0}
$$
\bea
\sigma_{ik}
&=&{\Gamma(m/2-1)\over 8\Gamma(m/2+2)} 
\Biggl\{
24\,{\mu_i^{-m/2+1}-\mu_k^{-m/2+1}\over (\mu_i-\mu_k)^3}
+8(m-4)\,{\mu_i^{-m/2}+\mu_k^{-m/2}\over  (\mu_i-\mu_k)^2 }
\nonumber\\[11pt]
&&
+{4(m-2)}\,{\mu_i^{-m/2}\over (\mu_i-\mu_k)^2}
+{m(m-2)}\,{\mu_i^{-m/2-1}\over  (\mu_i-\mu_k)}
\Biggr\},
\nonumber
\eea
\bea
<\Pi_i\tilde J_\alpha\Pi_k\tilde J_\beta>
&=&\sum_{1\le n,j\le s}{\Gamma(m/2)(2n+2j-2)!\over\Gamma(m/2+n+j-1)
2^{2n+2j-2}(n+j-1)!}\,
\nonumber\\[12pt]
&&\times 
g_{(\mu_1\mu_2}\cdots g_{\mu_{2n-3}\mu_{2n-2}}
g_{\nu_1\nu_2}\cdots g_{\nu_{2j-3}\nu_{2j-2}}g_{\gamma\delta)}\,
\nonumber\\[12pt]
&&\times
a^{(\mu_1\mu_2}\cdots a^{\mu_{2n-3}\mu_{2n-2}}a^\gamma{}_{\alpha}
a^{\nu_1\nu_2}\cdots a^{\nu_{2j-3}\nu_{2j-2}}a^{\delta)}{}_{\beta}\,.
\nonumber
\eea
\end{theorem}

\medskip
We shall prove this theorem in Section \ref{nine}.
Note that for Laplace type operators, when $a^{\mu\nu}=g^{\mu\nu}\II_{\cal V}$,
these formulas simplify considerably. We have then just one eigenvalue
$\mu_1=1$ with multiplicity $d_1=d=\dim({\cal V})$ and the projection
$\Pi_1=\II_{\cal V}$. Thus for Laplace type operators, 
$a_0=<\Pi_1>=\II_{\cal V}$,
and $\beta=-d/6$, and we recover the well known result 
\cite{gilkey95,avra91b}
%
%
$$
A_0=\vol(M),
$$
$$
A_1=\int_M d\vol(x) \left(\tr_{\cal V} q-{d\over 6}\,R\right)\,.
$$

In the next section we shall give a detailed description of our operator class.
We would like to stress from the beginning that the theory of NLT operators,
despite being a theory of second order operators, is closely related to the
theory of {\em higher-order Laplace-like operators}
(for a discussion, see \cite{avra98}); that is, operators with
principal part a power of the Laplacian.  In this sense, even the weighted form
Laplacian example mentioned above does not capture the full flavor of the
theory, since it (almost uniquely in this category) is accessible entirely
through second-order methods
(see \cite{gilkey91,gusynin99}).

To contrast with the previous result mentioned above, it is worth stressing 
once again that: 
\begin{itemize}
\item[i)] there are many areas in both physics and mathematics
where general second-order non-Laplace type operators arise naturally,
\item[ii)]  we are primarily interested not just in the weighted form
Laplacians but in the {\it general} NLT operators (differential forms being a
very particular case), 
\item[iii)] our approach to computation of heat kernel
asymptotics is completely different from that of the previous authors
\cite{gilkey91,gusynin99}. Our method enables us to compute  {\it explicitly}
not just the heat trace asymptotics but also the heat kernel and resolvent in
the leading order that describe the local off-diagonal behavior of the
resolvent and the heat kernel. To best of our knowledge, such formulas for
non-Laplace operators are presented here {\it for the first time}. 
\end{itemize}

Despite the importance of second-order operators
with non-Laplace principal part in gauge field theory and quantum gravity
(see, for example \cite{dewitt84,avra00}), their
study is still quite new, and the available methodology is still underdeveloped
in comparison with the Laplace type theory.  In this paper, and in the more
explicitly representation-theoretic treatment \cite{ab2}, we hope to lay the
groundwork for a systematic attack on the spectral asymptotics of this larger
class of operators.

\section{Non-Laplace Type Differential Operators}
\setcounter{equation}0

\subsection{General vector bundle setup}

Let $M$ be a smooth compact manifold without boundary of dimension $m$, 
equipped
with a (positive definite) Riemannian metric $g$.  Let $\cV$ be a smooth 
vector
bundle over $M$, with
${\rm End}(\cV)\cong \cV\otimes \cV^*$ the corresponding bundle
of endomorphisms.  Given any vector bundle $\cV$, we denote by
$C^\infty(M,\cV)$, or just $\ci(\cV)$, its space of smooth sections.  We assume
that the vector bundle $\cV$ is equipped with a Hermitian metric $H$.  This
naturally identifies the dual vector bundle $\cV^*$ with $\cV$, and defines a
natural $L^2$ inner product, using the invariant Riemannian measure $d\vol(x)$
on the manifold $M$.  The completion of $C^\infty(M,\cV)$ in this norm defines
the Hilbert space $L^2(M,\cV)$ of square integrable sections.

We denote by $TM$ and $T^*M$ the tangent and cotangent bundles of $M$.
We assume given a connection $\nabla^\cV$ on the vector bundle $\cV$.  The
covariant derivative on $\cV$ is then a map 
\be
\label{nv} 
\nabla^\cV: \
C^\infty(\cV)\to C^\infty(T^*M\otimes \cV) 
\ee
which we assume to be compatible with the Hermitian metric on the 
vector bundle
$\cV$, in the sense that $\N H=0$.  Here the connection is given its unique
natural extension to bundles in the tensor algebra over $\cV$ and $\cV^*$; in
particular, to the bundle $\cV^*\otimes \cV^*$ of which $H$ is a section.  In
fact, using the Levi-Civita connection $\N^{\LC}$ of the metric $g$ together
with $\N^\cV$, we naturally obtain connections on all bundles in the tensor
algebra over $\cV,\,\cV^*,\,TM,\,T^*M$; the resulting connection will 
usually be
denoted just by $\N$.  It will usually be clear which bundle's connection is
being referred to, from the nature of the section being acted upon.  
We denote
the curvature of $\N^\cV$ (a section of $T^*M\otimes T^*M\otimes \cV$) 
by ${\cal
R}$:
\be
[\N_\alpha\,,\N_\beta]\f=\cR_{\a\b}\f,\qquad
\f\in\ci(\cV).
\label{curv}
\ee

The formal adjoint of the covariant derivative of (\ref{nv}) is defined using
the Riemannian metric and the Hermitian structure on $\cV$:
\bes
\begin{array}{rl}
(\N^\cV)^*:\ci(T^*M\otimes \cV)&\to\ci(\cV), \\[10pt]
\f_\a\mapsto&-\N^\a\f_\a\,.
\end{array}
\ees

Let 
\bes
a\in\ci(TM\otimes TM)\otimes\endo(\cV)),
\ b\in\ci(TM\otimes \cV),\ q\in\ci(\endo(\cV)).
\ees
These sections define certain natural bundle maps by
contraction, which, for simplicity, we denote by the same letters:
\be
a: \ T^*M\otimes \cV\to TM\otimes \cV,
\label{mapa}
\ee
\bes
b: \ \cV\to TM\otimes \cV, 
\ees
\bes
q: \ \cV\to \cV.
\ees
Using these maps we can write the general second order operator
\be
F=\N^*(a\N)+b\N+q,\qquad{\rm where}\quad a^{\m\n}=a^{\n\m}.
\label{genlform-abs}
\ee
Any {\em formally self-adjoint} second order operator 
may thus be written
\be\label{genSA}
\N^*(a\N)+\frac12(b\N+(b\N)^*)+q,
\ee
where we may assume that each $a^{\m\n}$ is Hermitian.
(Here, if necessary, we clear the notation and redefine $q$ and 
$a^{\m\n}$.)
In abstract index notation, any formally self-adjoint 
second order operator may be written
\bea
F&=&-\N_\mu(a^{\mu\nu}\N_\nu)+b^\mu\N_\mu-\nabla_\mu (b^*)^\mu+q
\nonumber\\[10pt]
&=&
-a^{\mu\nu}\N_\mu \N_\nu+[b^\mu-(b^*)^\mu-a^{\nu\mu}{}_{;\nu}]\nabla_\mu
+q-(b^*)^\mu{}_{;\mu}\nonumber
\eea
Hereafter we denote by ";" the covariant derivative.
One may now restrict to the case when the endomorphism $b^*$ is anti-Hermitian,
since the Hermitian part of $b$ only 
contributes at order zero.
With this, we have redefined $b$ and $q$.
Thus,
henceforth,
\bes
a^{\mu\nu}=a^{\nu\mu}, \qquad b^*=-b.
\ees
The formal adjoint to the operator $F$ reads
\bes
F^*=\N^*(a^*\N)+b\N+(b\N)^*+q^*.
\ees
Hence, in addition to the other conditions posited so far,
for the operator $F$ to be formally self-adjoint the endomorphism
$q$ should be Hermitian.  To sum up, the general formally self-adjoint
second order operator is described by \nnn{genSA} with
\bes
a^{\m\n}=a^{\n\m},\ (a^{\m\n})^*=a^{\m\n},\ (b^\m)^*=-b^\m,\ q^*=q.
\ees

Let us consider the effect of a change of the connection, 
$\nabla\to\widetilde\nabla=\nabla+A$,
with a one form $A$ valued in $\End(\cal V)$. We have
\bes
\widetilde F=-\N_\mu a^{\mu\nu}\N_\nu
+(b^\mu-A_\nu a^{\nu\mu})\N_\mu
+\N_\mu(b^\mu-a^{\mu\nu}A_\nu)
+b^\mu A_\mu+A_\mu b^\mu
-A_\mu a^{\mu\nu} A_\nu+q.
\ees

In many cases (but not always!) it is possible to choose $A$ in such a way
that $A_\mu a^{\mu\nu}=b^\nu$. Then the first order part drops out.  
The point is whether the map $a$ (\ref{mapa}) is invertible, i.e.
whether there is a solution, $a^{-1}\in \ci(T^*M\otimes
T^*M)\otimes\endo(\cV)),$ to the equation
\be
a^{\mu\nu}a^{-1}{}_{\nu\lambda}=\delta^\mu_\lambda\II_{\cal V}.
\label{andg}
\ee
This can be put in another form. Let $e_i \in \ci(T^*M\otimes{\cal
V})$ be the basis in the space of one forms valued in ${\cal V}$ and
$e^*_i \in \ci(T^*M\otimes{\cal V}^*)$ be the adjoint basis in the
space of one forms valued in ${\cal V}^*$. Then the equation
(\ref{andg}) has a unique solution if
and only if the bilinear form $B_{ij}=<e^*_j, ae_i>$ is nondegenerate,
i.e. $\det B_{ij}\ne 0$.  If this condition is satisfied then one can
always redefine the connection in such a way, viz. $A_\mu=b^\nu
a^{-1}{}_{\nu\mu}$, that $A_\mu a^{\mu\nu}=b^\nu$ and the first order
terms are not present.  In this paper we assume that this is the case,
so that without loss of generality one can set the vector-endomorphism
$b$ to zero, $b=0$. Moreover, we will assume that the
tensor-endomorphism $a$ is parallel, $\nabla a=0$. Thus the operator
under consideration has the form 
\be
F=-a^{\mu\nu}\nabla_\mu\nabla_\nu+q,
\label{genlform}
\ee
where
\be
a^{\mu\nu}=a^{\nu\mu},\qquad a^{*}{}^{\mu\nu}=a^{\mu\nu},
\qquad \N a=0, \qquad q^*=q.
\label{gencond}
\ee

\subsection{Tensor-spinor bundles}

We now restrict attention to operators acting on {\it tensor-spinor bundles}.  
These bundles may be characterized as those appearing as 
direct summands of iterated 
tensor products of the cotangent and spinor bundles.  
Alternatively, they may
be described abstractly as bundles associated to representations of
$\orth(m)$, $\so(m)$, $\spin(m)$, or $\pin(m)$, depending on how much
structure we assume of our manifold.  These are extremely interesting
and important bundles, as they describe the fields in Euclidean
quantum field theory.  More general bundles appearing in field theory
are actually tensor products of these with auxiliary bundles, usually
carrying another (gauge) group structure.  
The connection on the tensor-spinor 
bundles is built in a canonical way from the Levi-Civita
connection and its curvature is:
$$
{\cal R}_{\mu\nu}=R^\alpha{}_{\beta\mu\nu}T^\beta{}_{\alpha},
$$
where $R^\alpha{}_{\beta\mu\nu}$ is the Riemann curvature tensor, and 
$T^\alpha{}_{\beta}$ is determined by the 
representation of ${\frak{so}}(m)$ which induces the bundle $\cV$.
$T^\beta{}_\alpha$ is a tensor-spinor 
constructed purely from 
Kronecker symbols, together with the fundamental tensor-spinor
$\g^\mu$ if spin structure is involved. 

We study in this paper a special class of second-order operators of the form
\nnn{genlform} with the coefficient $a$ built in a universal, polynomial way, 
using tensor product and contraction from the metric $g$ and its
inverse $g^*$, together with (if applicable) the volume form $E$ and/or
the fundamental tensor-spinor $\g$. Such a tensor-endomorphism $a$
is obviously
parallel. ($E$ is available given $\so(m)$ or $\spin(m)$
structure; $\g$ is available given $\spin(m)$ or $\pin(m)$ structure.)
We do not set any conditions on the endomorphism $q$,
except that it should be Hermitian. 
An important subclass of this class of operators
is the class of {\it natural} operators, when, in addition, 
$q$ is also built from the
geometric invariants 
only; i.e. from $g$, $g^*$, $E$, and $\gamma$, together with 
the Riemann curvature and its iterated covariant derivatives. 
By Weyl's invariant theory and dimensional analysis (i.e., a check of the
 homogeneity of each term under uniform dilation of the metric), it is clear
 that while $a$ must be built polynomially from $g$ and $g^*$, 
together with $E$ and/or $\g$
 if applicable,
the endomorphism $q$ must be a sum of terms linear in the curvature.
However, we do not need the additional assumption that $q$ is 
constructed from the curvature.
In general, it could be any smooth endomorphism. 
 
\section{Leading Symbol of an NLT Operator}
\setcounter{equation}0

Let us describe now more exactly the class of operators (\ref{genlform})
we are working with. We have assumed the operator $F$ to be 
{\it self-adjoint} leading to the conditions (\ref{gencond}).
Since we are going to study the heat kernel asymptotics of the operator $F$,
we now require in addition that the leading symbol of the operator $F$,
\bes
\sigma_2(F)(\x)=:A(\xi)=a^{\mu\nu}\xi_\mu\xi_\nu,
\qquad {\rm with}\ \xi\in T^*M,
\ees
be {\it positive definite}, i.e. we have
$$
\x\ne 0\Rightarrow A(\xi)\ \ {\rm Hermitian\ and\ positive\ definite\ on}\ \cV.
$$
In particular, $F$ is elliptic.
Positive definiteness implies that the roots of the characteristic polynomial
\be\label{CharPoly}
\chi_a(\x)(\l):=\det_\cV(A(\x)-\l)
\ee
are positive functions on $M$. 

\begin{remark}\label{constancy}
{\rm An important point is that the eigenvalues $\l_1\,,\ldots,\l_s$ and
their multiplicities $d_1\,,\ldots,d_s$
are independent of the point $x\in M$.  Indeed,
all the bundles under consideration are vector bundles associated to the
principal bundle $\cS$ of spin frames via a finite-dimensional representation 
$(\varphi,V)$ of 
$\spin(m)$: $\cV=\cS\times_\varphi V$.
Let $V_1$ be the defining representation of SO$(n)$.
We get the section $a$, and the endomorphisms $a^{\m\n}\x_\m\x_\n$ by
``promoting'' vectors
$\underline{a}\in V_1\ot V_1\ot V\ot V^*$ and 
$\underline{a}^{\m\n}\x_\m\x_\n\in V\ot V^*$ 
to sections of the associated bundles.  In particular,
the cited eigenvalues and multiplicities may be computed at the level
of the representation $(\varphi,V)$.}
\end{remark}

A useful way in which to perturb our operator $F$ is to let it run through
a one-parameter family $F(\e)$ for which
$$
a(\e)=a+\e b,
$$
where $b$ is a section of $TM\ot TM\ot\endo(\cV)$ which is built from
$g$ and $g^*$, and if applicable, $E$ and/or $\g$. (This $b$ is not to be
confused with the $b$ of section 2.1.)  To preserve 
formal self-adjointness, we need to assume that $b^{\m\n}\x_\m\x_\n$
is self-adjoint on $\cV$ for each $\x\in\ci(T^*M)$.
\bes
F=-(a^{\mu\nu}+\e b^{\m\n})\nabla_\mu\nabla_\nu+q.
\ees
For $\e$ in some interval about $0$, the symbol $a(\e)$
remains positive definite.  (This last statement does not even 
depend on the compactness of $M$; by an argument analogous to Remark
\ref{constancy}, this interval is independent of the point $x\in M$.)
In particular, we might take perturbations about an operator with
a leading symbol which is {\em factored}; i.e., one for which
$$
a^{\m\n}=g^{\m\n}c,
$$
where $c$ is a section of $\endo(\cV)$ which is built invariantly from 
$g$ and $g^*$, and if applicable, $E$ and/or $\g$. 
As a special case of this, we could take $c=\II_{\cV}\,$; that is,
perturb an operator of Laplace type in NLT directions.
In fact, in case $\cV$ is associated to an irreducible representation of
$\spin(m)$, the endomorphism $c$ {\em must} be a multiple of the
identity $\II_{\cV}\,$ by Schur's Lemma.

Given such a perturbation, one might hope that relevant spectral
quantities could be expanded in powers of (very small) $\e$, or at
least that one could work with the $\e$-variation of such quantities.

Now consider the symmetric $2n$-tensor quantity
\be\label{SymmA}
\tr_{\cal V} a^{(\mu_1\nu_1}\cdots a^{\mu_n\nu_n)}.
\ee
As usual, the parentheses denote complete symmetrization over all included
indices.
An index-free way of writing this is
\bes
\tr_{\cal V} \vee^n a,
\ees
where $\vee^n$ is the symmetrized tensor power defined by
\bes
\vee^n a\equiv \underbrace{a\vee\cdots\vee a}_{n}\,.
\ees

\begin{remark}\label{OrInsens}
We claim that 
\be
\tr_{\cal V} \vee^n a=a_{(n)}\vee^n g^*
\label{aton}
\ee
with some constants $a_{(n)}\,$.  
(Since $\N a$ and $\N g$ vanish, so must
$\N a_{(n)}\,$.)
\end{remark}

Indeed, this quantity is built polynomially from $g$ and 
$g^*$, and {\em a priori}
possibly $E$ and/or $\g$.  Since it is a tensor, $\g$ is not involved.
Since
$$
E^{\m_1\,\ldots\m_m}E_{\n_1\,\ldots\n_m}=\d^{[\m_1}{}_{[\n_1}\,\ldots
\d^{\m_m]}{}_{\n_m]}\,,
$$
where the brackets denote antisymmetrization over enclosed indices,
\nnn{SymmA} is affine linear in $E$; that is, it has the form
\bes 
\f^{(\m_1\n_1\,\ldots\m_n\n_n)}+
\y^{(\m_1\n_1\,\ldots\m_n\n_n)[\a_1\,\ldots\a_m]}E_{\a_1\,\ldots\a_m}.
\ees
The tensor $\psi$  is constructed purely from the metric 
$g^*$ and is symmetric
in the first $2n$ indices and antisymmetric in the last $m$ indices.  It is 
clear that there are no such tensors, so $\psi=0$.
Similarly, the only symmetric $2n$-tensor constructed from the metric
is $\vee^n g^*$, thus leading to (\ref{aton}).

Contracting (\ref{aton}) with $\x_{\m_1}\x_{\n_1}\ldots\x_{\m_n}\x_{\n_n}\,$,
we get
$$
\tr_{\cal V}A^n(\x)=a_{(n)}|\x|^{2n}\,.
$$
Furthermore, 
if we denote by $\tr_g$ the total trace of a symmetric $2n$-tensor 
$P$,
\be
\tr_g P\equiv g_{\mu_1\mu_2}\cdots g_{\mu_{2n-1}\mu_{2n}} 
P^{\mu_1\cdots\mu_{2n}}\,
\label{tottr}
\ee
then since
\bes
\tr_g \vee^n g^*=
{\Gamma(m/2+n)\over \Gamma(m/2)}{2^{2n}n!\over (2n)!} \,,
\ees
we have from (\ref{aton})
\bes
a_{(n)}={\Gamma(m/2)\over \Gamma(m/2+n)}{(2n)!\over 2^{2n}n!} 
\tr_{\cal V}\tr_g\vee^n a.
\ees
It is clear that for an operator of Laplace type, 
when $a=g^{*}\otimes {\rm id}_{\cal V}$,
each $a_{(n)}$ is just the fiber dimension $d$ of $\cV$.

Now let us consider the characteristic polynomial \nnn{CharPoly}
in more detail.  
Applying the above remarks to the coefficients of $\chi_a(\x)(\l)$, we 
find that $\chi_a(\x)$ depends on $\x$ only through $|\x|^2$.
As a result, the dependence of the eigenvalues $\l_i$ on $\x$ is
only through $|\xi|^2$.  Since $A(\x)$ is $2$-homogeneous in 
$\x$, the $\l_i$ must be also:
$$
\l_i(\x)=|\x|^2\m_i\,,
$$
for some positive real numbers $\m_1\,,\ldots,\m_s$ which are independent
of the point $(x,\x)\in T^*M$, and, in fact, independent of the specific
Riemannian manifold $(M,g)$, depending only on the representation
$(\varphi,V)$ and the vector 
$\underline{a}\in V_1\ot V_1\ot V\ot V^*$.
Computing the trace $\tr_{\cal V}A(\x)^n$ we get a sequence of equations,
\be\label{MuDandA}
\sum_{i=1}^sd_i\m_i^n=a_{(n)}\,,\qquad n\in\NN,
\ee
relating the eigenvalues, multiplicities, and the quantities $a_{(n)}\,$.
When $n=0$, \nnn{MuDandA} is just $a_{(0)}=d$; this was immediate from
the definition of $a_{(n)}\,$.  

Let $\Pi_i(\xi)$ be the orthogonal projection onto the $\l_i\,$-eigenspace.
The $\P_i$ satisfy the conditions
\bea
\Pi_i^2&=&\Pi_i,\nonumber \\
\Pi_i \Pi_k&=&0 \qquad (i\ne k),\nonumber \\
\sum_{i=1}^s \Pi_i&=&\II_{\cal V},\nonumber \\
\tr_{\cal V} \Pi_i&=&d_i.
\label{Qxxx}
\eea
In contrast to the eigenvalues, the projections 
depend on the {\em direction} $\xi/|\xi|$ of $\xi$, rather than on
the magnitude $|\x|$.
In other words, they are $0$-homogeneous in $\xi$.  Furthermore, they
are polynomial in $\xi/|\xi|$:
\be\label{polyns}
\Pi_i(\xi)=\sum_{n=0}^{2p} P_{i(n)}(\xi),
\ee
for some $p$, 
where
\bes
P_{i(n)}(\xi)={1\over |\xi|^{n}}\xi_{\mu_1}\cdots\xi_{\mu_n}
\Pi_{i(n)}^{\mu_1\cdots\mu_n}.
\ees
Here the $\Pi_{i(n)}$ 
are some $\End(V)$-valued symmetric $n$-tensors
that do not depend on $\xi$. 

There is, however, quite a bit of ambiguity in the definition 
\nnn{polyns} of the homogeneous polynomials
$P_{i(n)}(\xi)\,$, since 
multiplication of an $n$-homogeneous polynomial $q(\x)$ by $|\x|^2$ produces
an $(n+2)$-homogeneous polynomial $\tilde q(\x)$, without changing
the associated $0$-homogeneous function $q(\x/|\x|)$.
We can remove the ambiguity by requiring that $P_{i(n)}$ have no
$|\x|^2$ factor.  This is equivalent to requiring that 
$P_{i(n)}(\xi)$ be a {\em harmonic} polynomial in $\x$, 
which in turn is equivalent
to requiring that its restriction to the unit $\x$-sphere is an
$\nth\,$-order ($\endo(\cV)$-valued) 
spherical harmonic.  Yet another equivalent formulation
is to require that 
$\Pi_{i(n)}$ is trace free in all its indices.
At any rate, with this requirement, we have uniquely defined quantities
$\Pi_{i(n)}\,$.
 
Note that the explicit formula
$$
\P_i=
\frac{(A-\l_1)\ldots(A-\l_{i-1})(A-\l_{i+1})\ldots(A-\l_s)}
{(\l_i-\l_1)\ldots(\l_i-\l_{i-1})(\l_i-\l_{i+1})\ldots(\l_i-\l_s)}
$$
exhibits each projection as a polynomial of degree $2(s-1)$.
By homogeneity, 
$$
\P_i(\x)=
\frac{(A({\x/|\x|})-\m_1)\ldots(A({\x/|\x|})-\m_{i-1})(A({\x/|\x|}) 
-\m_{i+1})\ldots(A({\x/|\x|})-\m_s)}
{(\m_i-\m_1)\ldots(\m_i-\m_{i-1})(\m_i-\m_{i+1})\ldots(\m_i-\m_s)}.
$$
We can take this expression and expand as a product of 
homogeneous terms:
\be
\P_i(\x)=\sum_{k=1}^{s}c_{ik}A^{k-1}\left({\x/|\x|}\right)
=\sum_{k=1}^s c_{ik}a^{(\mu_1\mu_2}\cdots a^{\mu_{2k-3}\mu_{2k-2})}
{\xi_{\mu_1}\cdots\xi_{\mu_{2k-2}}\over |\xi|^{2k-2}},
\label{PiToAk}
\ee
where $c_{ik}$ are numerical constants depending only on the $\mu_j$.
Since $A(\xi)$ is $2$-homogeneous it follows that all homogeneity orders are 
even, i.e. $P_{i(2n+1)}(\xi)=\P_{i(2n+1)}=0$, and therefore,
\be
\Pi_i(\xi)=\sum_{n=0}^p P_{i(2n)}(\xi).
\label{321}
\ee
In turn, by writing $A(\xi/|\xi|)=\sum_{i=0}^s \mu_i \Pi_i(\xi)$
we compute powers of $A$
in terms of projections:
$$
A^k(\xi/|\xi|)=\sum_{i=1}^s \mu^k_i \Pi_i(\xi).
$$
Substituting this into \nnn{PiToAk}, we get
$$
\sum_{k=1}^sc_{ik}\m_j^{k-1}=\d_{ij}\,.
$$ 
In other words, the matrix of coefficients $\cC:=(c_{ik})$
is inverse to the (Vandermonde) matrix of powers
$\cM=(\kappa_{kj}):=(\m_j^{k-1})$:
\be
(c_{ik}):=(\kappa_{kj})^{-1}\;.
\label{cik}
\ee

Though the number $s$ depends on the particular leading symbol $a$,
there is an upper bound for $s$ which depends only on the
representation $(\f,V)$ to which the bundle $\cV$ is associated.  This
is described in detail in \cite{ab2}.  In particular, in case $(\f,V)$
is irreducible, the algebra generated by restrictions to the unit
$\x$-sphere of equivariant leading symbols of all orders is
commutative, and thus simultaneously diagonalizable.  The resulting
projections diagonalize equivariant leading symbols of any (not just
second) order.  The number of projections, i.e.\ the dimension of the
algebra just described, may be described in terms of
representation-theoretic parameters.
For reducible $(\f,V)$, similar considerations are valid, but the
algebra is not commutative.

Now write the leading symbol 
in terms of projections
\be
A(\x)=|\xi|^2\sum_{i=1}^s\m_i\P_i(\x)
=|\xi|^2\sum_{i=1}^s\sum_{n=0}^p\m_i P_{i(2n)}(\x).
\label{AtoPi}
\ee
Since $A(\x)$ is $2$-homogeneous, it is a sum of $\endo(\cV)$-valued
spherical harmonics in $\x$ of degrees $2$ and $0$.  Thus the 
spherical harmonics on the right of all other orders vanish:
\bes
\sum_{i=1}^s\mu_i P_{i(2n)}(\xi)
=\sum_{i=1}^s\mu_i \Pi_{i(2n)}=0,\qquad n\ne 0, 1.
\ees

Furthermore, it is clear that 
the decomposition of $A(\x)$ into $n=2$ and $n=0$ contributions
comes upon taking
$$
A(\xi)=a^{\m\n}\x_\m\x_\n=|\x|^2b_0+b_2^{\m\n}\x_\m\x_\n\,,
$$
where $b_2^{\m\n}\x_\m\x_\n$ is a second-order spherical harmonic in 
$\x$; i.e., $b_2^{\m\n}$ is trace free in its two indices.  This gives
\bes
a^{\mu\nu}=g^{\mu\nu}b_0 + b_2^{\mu\nu},
\ees
where
\bes
b_0={1\over m}g_{\mu\nu}a^{\mu\nu}
=\sum_{i=1}^s\mu_i\Pi_{i(0)}, \qquad
b_2^{\mu\nu}=\sum_{i=1}^s\mu_i\Pi_{i(2)}^{\mu\nu}.
\ees

\section{Symmetric Two-Tensors}
\setcounter{equation}0

It is instructive at this point to consider two examples:
the bundle $\cS^2$ of symmetric two-tensors, and the subbundle 
$\cS^2_0$ of trace-free symmetric two-tensors.
In particular, in these examples, one begins to glimpse the differences
and relations between the cases of reducible and irreducible $\cV$.
We may compute with tensors valued in either complex or real tensor
bundles;
in fact, most of the following  
discussion 
holds in either setting.
But for the sake of definiteness, and with a view toward applications,
let us assume that all our tensors are real.

First note that a basis of the 
$0$-homogeneous symbols in our class is given by
\bes
(X_1\varphi)_{\alpha\beta}=g_{\alpha\beta}\varphi^\mu{}_\mu\,,
\ees
\bes
(X_2\varphi)_{\alpha\beta}
={1\over |\xi|^2}\xi^\mu\xi_{(\alpha}\varphi_{\beta)\mu}\,,
\ees
\bes
(X_3\varphi)_{\alpha\beta}
={1\over |\xi|^2}\xi_\alpha\xi_\beta\varphi^\mu{}_\mu\,,
\ees
\bes
(X_4\varphi)_{\alpha\beta}
={1\over |\xi|^2}g_{\alpha\beta}\xi^\mu\xi^\nu\varphi_{\mu\nu}\,,
\ees
\bes
(X_5\varphi)_{\alpha\beta}
={1\over |\xi|^4}\xi_\alpha\xi_\beta\xi^\mu\xi^\nu\varphi_{\mu\nu}\,.
\ees
Note that in the inner product $(\varphi,\psi)=\varphi_{\mu\nu}\psi^{\mu\nu}$,
we have
$$
X_1^*=X_1,\qquad X_2^*=X_2, \qquad X_3^*=X_4, \qquad X_5^*=X_5.
$$
In addition, the traces of these endomorphisms are
\bes
\tr_{\cS^2}\II_{\cS^2}={m(m-1)\over 2}\,,
\ees
\bes
\tr_{\cS^2}X_1=m,
\ees
\bes
\tr_{\cS^2}X_2=m+1,
\ees
\bes
\tr_{\cS^2}X_3
=\tr_{\cS^2}X_4
=\tr_{\cS^2}X_5=1.
\ees
The multiplication table of the algebra generated by 
$\II_{\cV}$ and the $X_i$ is shown in Table 1.
\begin{table}[ht]
\begin{center}
\begin{tabular}{|c||c|c|c|c|c|}
\hline
$\to$	&$X_1$&$X_2$&$X_3$&$X_4$&$X_5$\\
\hline
\hline
$X_1$	&$mX_1$ &$X_4$ &$X_1$ &$mX_4$ &$X_4$\\
\hline
$X_2$	&$X_3$  &${1\over 2}(X_2+X_5)$ &$X_3$ &$X_5$ &$X_5$\\
\hline
$X_3$	&$mX_3$ &$X_5$ &$X_3$ &$mX_5$ &$X_5$\\
\hline
$X_4$	&$X_1$  &$X_4$ &$X_1$ &$X_4$ &$X_4$\\
\hline
$X_5$	&$X_3$  &$X_5$ &$X_3$ &$X_5$ &$X_5$\\
\hline
\end{tabular}
\end{center}
\caption{Multiplication table}
\end{table}
The arrow indicates that the left factor is to be found in the leftmost
column.  For
example, $X_3X_4=mX_5\,$.

The leading symbol of any even-order operator has
the form
\bes
A(\xi)=
|\xi|^{2p}\left(\alpha_0\,\II_{\cS^2}
+\alpha_1\, X_1
+2\alpha_2\, X_2
+\alpha_3\, X_3 
+\alpha_4\, X_4+\alpha_5\,X_5\right),
\ees
for some $p$ and 
some numerical parameters $\alpha_i$. 
If $A$ is the leading symbol of a second-order operator,
then $\alpha_5$ vanishes.
If in addition $A$ is self-adjoint,
then $\alpha_3=\alpha_4\,$.
Thus for a second-order self-adjoint leading symbol $A$,
\be\label{redu}
A(\xi)=
|\xi|^2\left(\alpha_0\,\II_{\cS^2}
+\alpha_1\, X_1
+2\alpha_2\, X_2
+\alpha_3(X_3+X_4) 
\right).
\ee
If we further require that $A$ be positive definite, we get 
additional inequality constraints on the $\a_i\,,$;
these are
discussed below.

\subsection{Trace-free symmetric two-tensors}

Now consider the bundle $\cS^2_0$ of trace-free symmetric two-tensors.
The symbol 
\bes
P:=\II_{\cS}-{1\over m}X_1
\ees
is the self-adjoint projection onto 
${\cal S}_0^2$.
Thus we may define a spanning set of symbols $Y_i$ 
on ${\cal S}_0^2$ by
\bes
Y_i=PX_iP.
\ees
According to Table 1,
$$
Y_1=Y_3=Y_4=0,
$$
and
\bes
Y_2=X_2-{1\over m}(X_3+X_4)+{1\over m^2}X_1,
\ees
\bes
Y_5=X_5-{1\over m}(X_3+X_4)+{1\over m^2}X_1\,.
\ees
Clearly both $Y_2$ and $Y_5$ are self-adjoint.
Together with the identity symbol $\II_{\cS^2_0}\,$, 
the symbols $Y_2$ and $Y_5$ form a basis of the symbol
algebra on $\cS^2_0\,$. 
For purposes of explicit computation, it is useful to note that
\be
PX_1=X_1P=PX_4=X_3P=0.
\label{PX1}
\ee
We compute that
\be
Y_2Y_5=Y_5Y_2=Y_5^2={m-1\over m}Y_5\,,
\label{Y52}
\ee
\be\label{y2sq}
Y_{2}^2=\dfrac12Y_{2}+\dfrac{m-2}{2m}Y_{5}\,.
\ee
This points up a very important and useful fact:
that the symbol algebra over $\cS^2_0$ is {\em commutative}.
In fact, as shown in \cite{ab2}, this is a general feature
of the case of an irreducible bundle.  In the case of a bundle, like
$\cS^2$, that is reducible under its structure group, the symbol
algebra may be noncommutative.  

The commutativity of the symbol algebra over $\cS^2_0$ significantly
simplifies the computation of the projections.  First note that all
symbols will be simultaneously diagonalizable, so the discussion on
this level is independent of the particular symbol $A$.
{}From (\ref{Y52}) we immediately
obtain one projection, namely
\be\label{insproj}
\Pi_{3}={m\over m-1}Y_5\,.
\ee
Since $\II_{\cS^2_0}\,$, $Y_2\,$, 
and $Y_5$ form a basis of the symbol algebra, (\ref{y2sq}) shows that
$\II_{\cS^2_0}\,$, $Y_2\,$, and $Y_2^2$ are also a basis.
Note that
\bes
Y_{2}^3=\dfrac{3m-2}{2m}Y_{2}^2-\dfrac{m-1}{2m}Y_{2}\,,
\ees
and
\bes
Y_{2}^4=\dfrac{7m^2-10m+4}{4m^2}Y_{2}^2-\dfrac{(3m-2)(m-1)}{4m^2}Y_{2}.
\ees
If $\P:=a\II_{\cS^2_0}+bY_{2}+cY_{2}^2$, we thus get
$$
\begin{array}{l}
\P^2=a^2\II_{\cS^2_0}
+\left\{2ab-\dfrac{bc(m-1)}{m}-\dfrac{c^2(3m-2)(m-1)}{4m^2}\right\}Y_{2} 
\\[11pt]
\qquad+\left\{2ac+b^2+\dfrac{bc(3m-2)}{m}
+\dfrac{c^2(7m^2-10m+4)}{4m^2}\right\}Y_{2}^2.
\end{array}
$$
There are $8=2^3$ solutions $(a,b,c)$ of the projection equation
$\P^2=\P$.  Two of these are
$$
(a_2\,,b_2\,,c_2):=\left(0,\dfrac{4(m-1)}{m-2}\,,-\dfrac{4m}{m-2}\right)
$$
and 
$$
(a_3\,,b_3\,,c_3):=\left(0,-\dfrac{m^2}{(m-1)(m-2)}\,,\dfrac{2m^2}{(m-1)(m-2)}
\right)\,;
$$
these in fact correspond to {\em fundamental} projections $\P_2$ and $\P_3\,$.
($\P_3$ is the same as the projection found by inspection in
(\ref{insproj}).)

The remaining solutions correspond to the third fundamental projection
$\P_1:=\II_{\cS^2_0}-\P_2-\P_3\,$, and to 
$0$, $\II_{\cS^2_0}\,$, $\P_2+\P_3\,$, $\P_1+\P_3\,$, and $\P_1+\P_2\,$.
The fundamental projections have particularly simple expressions in
terms of $Y_{2}$ and $Y_{5}$:
\bea
\label{ProjT}
\P_2&=&\dfrac4{m-2}(-mY_{2}^2+(m-1)Y_{2})=
2(Y_{2}-Y_{5}) \\[10pt]
\P_3&=&\dfrac{m^2}{(m-1)(m-2)}(2Y_{2}^2-Y_{2})=
\dfrac{m}{m-1}Y_{5} \\[10pt]
\P_1&=&\II_{\cS^2_0}-2Y_{2}+\dfrac{m-2}{m-1}Y_{5}
\eea
The general self-adjoint 
second-order symbol $A_0$ on $\cS^2_0$ may be viewed
as the compression $PAP$ of a second-order symbol $A$ on $\cS^2$, which,
in view of (\ref{redu}), is
\bea
A_0(\xi)&=&|\xi|^2\left(\alpha_0\II_{\cS^2_0}+2\alpha_2 Y_2\right)
\nonumber\\[10pt]
&=&|\xi|^2\left(\m_1\P_1+\m_2\P_2+\m_3\P_3\right)
\nonumber\\[10pt]
&=&|\xi|^2\Big\{\m_1\II_{\cS^2_0}+2(\mu_2-\mu_1)Y_2
\nonumber\\[10pt]
&&
+{1\over m-1}\left[(m-2)\mu_1-2(m-1)\mu_2+m\mu_3\right]Y_5\Big\}.
\nonumber
\eea
{}From this we obtain the eigenvalues 
in terms of the parameters $\alpha_0$ and $\alpha_2\,$:
\bes
\mu_1=\alpha_0, \qquad \mu_2=\alpha_0+\alpha_2,
\qquad
\mu_3=\alpha_0+{2(m-1)\over m}\alpha_2\,.
\ees
Thus, self-adjoint positive definite 
second-order symbols are in one-to-one correspondence with 
choices of $(\alpha_0\,,\alpha_2)$
for which
\bes
\alpha_0\,,\;\alpha_0+\alpha_2\,,\;
m\alpha_0+2(m-1)\alpha_2\,\;>0.
\ees
Note that for $m\ge 2$, the first and third conditions
together imply the second one.

Our projections have the following interpretation.  After choosing
the distinguished direction $\x$, one can distinguish three subspaces
of the trace-free symmetric two-tensors.  First, there are tensors
in the direction of the trace-free part $(\x\ot\x)_0$ 
of $\x\ot\x$.  This is clearly
the range of $\P_3\,$, since $Y_{5}\f$ is a scalar multiple of $(\x\ot\x)_0\,$.
Next, there is the subspace consisting of tensors $\x\vee\z$, where
$\z\perp\x$.  This is the range of $\P_2\,$, since for
$$
\z_\b:={1\over |\xi|}\x^\l\f_{\b\l}
-{1\over |\xi|^3}\x_\b\x^\l\x^\m\f_{\l\m}\,,
$$
we have 
$$
(Y_{2}-Y_{5})\f=\x\vee\z,\qquad\z^\l\x_\l=0.
$$
The remaining subspace is the range of $\P_1\,$; it may be described
as the space generated by tensors $(\z\vee\theta)_0\,$, where
$\z\perp\x\perp\theta$.
The traces of the projections are thus
$$
\tr_\cV\P_1=\dfrac{(m+1)(m-2)}2\,,\qquad
\tr_\cV\P_2=m-1,\qquad\tr_\cV\P_3=1.
$$

\subsection{Symmetric two-tensors (not necessarily trace-free)}

To leave the irreducible setting, consider the bundle $\cS^2$ of symmetric
two-tensors (unrestricted as to trace); this is equivariantly 
isomorphic to the direct sum of the trace-free symmetric two-tensors
and the scalars.  In fact, the direct sum decomposition 
is implemented by the projections
$P$ and $\II_{\cS^2}-P=(1/m)X_1\,$.
By noting that $X_1=m(\II_{\cS^2}-P)$ and 
using Table 1, we obtain in addition to (\ref{PX1})
\bea
&&PX_2(\II_{\cal V}-P)={1\over m}\left(X_3-{1\over m}X_1\right),
\nonumber\\[10pt]
&&(\II_{\cal V}-P)X_2P={1\over m}\left(X_4-{1\over m}X_1\right),\qquad 
\nonumber\\[10pt]
&&PX_3(\II_{\cal V}-P)=X_3-{1\over m}X_1,\qquad \nonumber\\[10pt]
&&(\II_{\cal V}-P)X_4P=X_4-{1\over m}X_1\,,\nonumber\\[10pt]
&&(\II_{\cal V}-P)X_2(\II_{\cal V}-P)={1\over m}(\II_{\cal V}-P), \qquad 
\nonumber\\[10pt]
&&(\II_{\cal V}-P)X_3(\II_{\cal V}-P)=(\II_{\cal V}-P)X_4(\II_{\cal V}-P)
=(\II_{\cal V}-P).
\label{PXT}
\eea
Let $\Pi_1$, $\Pi_2$ and $\Pi_3$ be the fundamental projections
defined by (\ref{ProjT}). Denoting
\bes
T={1\over\sqrt{m-1}}\left(X_3-{1\over m}X_1\right),\qquad
T^*={1\over\sqrt{m-1}}\left(X_4-{1\over m}X_1\right)
\ees
and using the equations (\ref{PXT}), we obtain the decomposition 
$$
A(\xi):=|\xi|^2\left[\mu_1\Pi_1
+\mu_2\Pi_2
+\mu_3\Pi_3
+\kappa\left(T+T^*\right)
+q(\II_{\cS^2}-P)\right]\,,
$$
where $\kappa$ and $q$ are real constants defined by
$$
\kappa=\sqrt{m-1}\left({2\over m}\alpha_2+\alpha_3\right),
$$ 
\bes
q=\alpha_0+m\alpha_1+{2\over m}\alpha_2
+2\alpha_3\,.
\ees
Note the useful relations
\bes
TP=T\Pi_1=T\Pi_2=T\Pi_3=0,
\ees
\bes
PT^*=\Pi_1T^*=\Pi_2T^*=\Pi_3T^*=0.
\ees
$$
T^2=(T^*)^2=0
$$
$$
T^*T=\II_{\cal V}-P,\qquad
TT^*=\Pi_3\,.
$$
\be
\Pi_3T=T,\qquad T^*\Pi_3=T^*.
\label{PiT}
\ee

Observe also
the obvious {\it non-commutativity}; the projections will now depend on
the particular leading symbol we are studying.  
We see that 
for our new symbol on $\cS^2$, the first two projections
$\P_1\,$, $\P_2$ 
computed above are untouched, and there are two more projections 
$Z_3$, $Z_4$ onto
one-dimensional subspaces.  
$\P_1$ and $\P_2$ are independent of the particular symbol we are 
diagonalizing, while
and $Z_3$ and $Z_4$ depend on it.
$Z_3$ and $Z_4$ take the form
\bes
Z=a\Pi_3+b(T+T^*)+c(\II_{\cal V}-P)
\ees
By (\ref{PiT}), we have
\bes
Z^2=(a^2+b^2)\Pi_3+b(a+c)(T+T^*)+(b^2+c^2)(\II_{\cS^2}-P).
\ees
The projection equation $Z^2=Z$ gives
$$
(2a-1)^2+(2b)^2=(2c-1)^2+(2b)^2=1,\qquad b(a-c+1)=0.
$$
Aside from the solutions
\be\label{degensol}
(a,b,c)=(0,0,0),\ (1,0,1),
\ee
all solutions have the form
\bes
a={1\over 2}(1+\cos\theta), \qquad 
b={1\over 2}\sin\theta,\qquad
c={1\over 2}(1-\cos\theta),
\ees
where $\theta$ is an arbitrary real parameter; conversely, all 
choices of $\theta$ give solutions.
The solutions (\ref{degensol}) give $0$- and $2$-dimensional projections,
so can be discarded from the present point of view, where we seek 
complementary one-dimensional projections.
If we define
\bea
(a_3, b_3, c_3):&=&\left({1\over 2}(1+\cos\theta),  \ {1\over 2}\sin\theta, \ 
{1\over 2}(1-\cos\theta)\right), 
\nonumber\\[10pt]
(a_4, b_4, c_4):&=&\left({1\over 2}(1-\cos\theta),  \ -{1\over 2}\sin\theta, \ 
{1\over 2}(1+\cos\theta)\right),
\nonumber
\eea
we get a set of complementary projections $Z_3$ and $Z_4\,$; that is,
we have
$Z_3Z_4=Z_4Z_3=0$.

We still need to find a value of $\theta$ adapted to our given
symbol $A$.
Denote by $\nu_3$ and $\nu_4$ the eigenvalues 
of $A$ in the ranges of  
$Z_3$ and $Z_4$ respectively:
$$
\mu_3\Pi_3
+\kappa\left(T+T^*\right)
+q(\II_{\cal V}-P)=
\nu_3 Z_3+\nu_4 Z_4.
$$
We find that
\bes
\begin{array}{l}
(\nu_3+\nu_4)+(\nu_3-\nu_4)\cos\theta=2\mu_3,\\
(\nu_3-\nu_4)\sin\theta=2\kappa,\\
(\nu_3+\nu_4)-(\nu_3-\nu_4)\cos\theta=2q.
\end{array}
\ees
{}From these equations we first determine the eigenvalues
$$
\nu_3=\rho+\omega, \qquad
\nu_4=\rho-\omega,
$$
where
\bes
\rho={1\over 2}(\mu_3+q)=\alpha_0+{m\over 2}\alpha_1+\alpha_2+\alpha_3,
\ees
\bea
\omega^2&=&\kappa^2+\left({\mu_3-q\over 2}\right)^2
\nonumber\\[10pt]
&=&{1\over 4m^2}\left[-m^2\alpha_1+2(m-2)\alpha_2-2m\alpha_3\right]^2
+4(m-1)(2\alpha_2+m\alpha_3)^2\,.
\nonumber
\eea
The positivity of the leading symbol is translated now into the condition
$\omega^2<\rho^2$, or $\kappa^2<q\mu_3$. So, in addition to $\alpha_0>0$
and $\alpha_0+\alpha_2>0$ we have
$$
(m-1)(2\alpha_2+m\alpha_3)^2<(m\alpha_0+m^2\alpha_1+2\alpha_2+2m\alpha_3)
\left[m\alpha_0+2(m-1)\alpha_2\right].
$$

The parameter $\theta$ is now determined by 
\bea
\cos\theta &=&{\mu_3-q\over 2\omega}={1\over 2m\omega}\left[-m^2\alpha_1
+2(m-2)\alpha_2-2m\alpha_3\right], \qquad
\nonumber\\[10pt]
\sin\theta &=&{\kappa\over\omega}={\sqrt{m-1}\over m\omega}\left(2\alpha_2
+m\alpha_3\right).
\nonumber
\eea
Therefore, the projections $Z_3$ and $Z_4$ have the form
$$
Z_{3,4}={1\over 2}\left(1\pm{\mu_3-q\over 2\omega}\right)\Pi_3
\pm{\kappa\over 2\omega}(T+T^*)
+{1\over 2}\left(1\mp{\mu_3-q\over 2\omega}\right)(\II_{\cal V}-P),
$$
and indeed depend on the symbol $A$.

\section{The resolvent and the heat kernel}
\setcounter{equation}0

Let $F$ be a self-adjoint 
second-order partial differential operator on a compact manifold with
positive definite leading symbol.  (In particular, $F$ is elliptic.)  
Then $F$ has discrete real eigenvalue spectrum
which is bounded below by some (possibly negative) real number $c$.
If $\l$ is a complex number with ${\rm Re}\,\l<c$, then 
the {\em resolvent} $(F-\l\II)^{-1}$ is well defined: if
$\{(\l_j\,,\f_j)\}$ is a spectral resolution, with the $\f_j$ forming
a complete orthonormal set in $L^2(\cV)$, then
$$
(F-\l\II)^{-1}\f_j=(\l_j-\l)^{-1}\f_j\,.
$$
The resolvent is a bounded operator on $L^2(\cV)$, and in fact is
a compact operator by the Rellich Lemma, since it carries
$L^2(\cV)$ to the Sobolev section space $L^2_2(\cV)$
continuously.
The resolvent kernel is a
section of the external tensor product of the vector bundles $\cV$ and
$\cV^*$ over the product manifold $M\times M$, and satisfies the
equation 
\bes 
(F-\lambda\II)G(\lambda|x,y)=\delta(x,y)
\ees
where $\delta(x,y)$ is the 
Dirac distribution (which in turn is the kernel function 
of the identity operator).
Here and below, all differential operators will act in the {\it first}
($x$, as opposed to $y$) argument of any kernel 
functions to which they are applied.
The resolvent is well-defined as long as the null space of $F-\l\II$
vanishes; i.e., as long as $\l$ is not one of the $\l_j\,$.  As a 
consequence of self-adjointness, we
have
\bes
{G^{\dag}(\lambda|x,y)}=G(\bar\lambda|y,x).
\ees

Similarly, for $t>0$ the heat 
operator $U(t)=\exp(-tF):\ L^2(M,V)\to L^2(M,V)$
is well defined. 
$U(t)$ is a {\em smoothing} operator; that is, it carries $L^2$
sections to sections in $\cap_{k\in\bbR}L^2_k=\ci$.
The kernel function of this operator, called the {\em heat kernel}, 
satisfies the equation
\bes
(\partial_t+F)U(t|x,y)=0
\ees
with the initial condition
\bes
U(0^+|x,y)=\delta(x,y),
\ees
and the self-adjointness condition
\bes
{U^{\dag}(t|x,y)}=U(t|y,x).
\ees

As is well known \cite{gilkey95}, the heat 
kernel and the resolvent kernel are related
by the Laplace transform:
\bes
G(\lambda)=\int\limits_0^\infty dt\, e^{t\lambda}\,U(t),
\ees
\bes
U(t)={1\over 2\pi i}\int\limits_{c-i\infty}^{c+i\infty}d\lambda\, 
e^{-t\lambda}\,G(\lambda).
\ees
It is also 
well known 
\cite{gilkey95} that the heat kernel $U(t|x,y)$ 
is a smooth function near the diagonal $\{x=y\}$
of $M\times M$,
with the diagonal values integrating to the functional trace:
\bes
\Tr_{L^2}\exp(-tF)=\int_M d\vol(x)\tr_{\cal V} U(t|x,x).
\ees
Moreover, there is an asymptotic expansion of the heat kernel
as $t\to 0^+$,
\bes
\Tr_{L^2}\exp(-tF)\sim 
(4\pi t)^{-m/2}\sum\limits_{k\ge 0}{(-t)^{k}\over k!}A_k,
\ees
and a corresponding expansion of the resolvent as $\lambda\to-\infty$:
\bes
\Tr_{L^2}\partial_\lambda^n G(\lambda)
\sim (4\pi)^{-m/2}\sum\limits_{k\ge 0}{(-1)^k\over k!}\Gamma[(k-m)/2+n+1]
(-\lambda)^{m/2- k-n-1}A_k,
\ees
for $n\ge m/2$. 
Here $A_k$ are the {\it global}, or {\it integrated} heat coefficients,
sometimes called the Minakshisundaram-Pleijel coefficients.

There is additional information in the 
{\em local} asymptotic expansion of the heat 
kernel diagonal,
\be
U^{\rm diag}(t):=
U(t|x,x)\sim (4\pi t)^{-m/2}\sum\limits_{k\ge 0}{(-t)^{k}\over k!}a_k\,.
\label{566xxx}
\ee
The {\em local} heat coefficients
$a_k$ integrate to the global ones:
\be
A_{k}=\int\limits_M d\vol(x)\,\tr_{\cal V} a_k.
\label{Ak}
\ee
One can, in fact, get access to the local heat coefficients via a 
functional trace, by taking advantage of the principle that 
a function (or distribution) is determined by its integral against
an arbitrary test section $f\in C^\infty(\End(\cal V))$:
\bes
\Tr_{L^2}\,f\exp(-tF)\sim
(4\pi t)^{-m/2}\sum\limits_{k\ge 0}{(-t)^{k}\over k!}A_k(f,F).
\ees
We have 
\bes
A_k=A_k(1,F),
\qquad A_k(f,F)=\int\limits_Md\vol(x)\,f\,\tr_{\cal V}\,a_k\,.
\ees
The local heat
coefficients $a_k$ have been calculated for Laplace type operators up to
$a_4$ \cite{avra91b}.  For non-Laplace type operators, some of them are
known only in the very specific case of differential form
bundles \cite{gilk-bransMN}.

We shall calculate below the coefficients 
$A_0$ and $A_1$ for NLT operators in terms of the projections introduced
in the previous sections.
Our tactic
will be to construct an approximation
to the heat kernel $U(t|x,y)$, that is a  
{\it parametrix}.  The important information in the parametrix 
for small $t$ is 
carried by its values near the diagonal,
since the heat kernel
vanishes to order $\infty$ off the diagonal as $t\to 0^+$.
Since the heat
and resolvent kernels are related by the Laplace transform, this 
is equivalent to studying an approximation
to the resolvent kernel $G(\lambda|x,y)$ near the diagonal for 
large negative ${\rm Re}\,\lambda$. 

Let us stress here that our purpose is not to provide a rigorous
construction of the resolvent with estimates; for this we rely on the
standard references \cite{gilkey95}.  Rather, given that the existence of
resolvent and heat parametrices is known, our aim is to compute
various aspects of it; in particular, information on leading order
terms sufficient to determine some of the heat kernel coefficients
$A_{k}$.
 
We shall employ the standard scaling device
for the 
resolvent $G(\lambda|x,y)$ and heat kernel $U(t|x,y)$ when
$x\to y$, $\lambda\to -\infty$ and $t\to 0$. 
This means that one introduces a small expansion parameter 
$\varepsilon$ reflecting the fact that 
the points $x$ and $y$ are close to each other, the parameter
$t$ is small, and the parameter ${\rm Re}\,\lambda$ is negative and large. 
This can be done by fixing a point $x'$, choosing normal 
coordinates at $x'$ (with $g_{\mu\nu}(x')=\delta_{\mu\nu}$), 
scaling according to
\bea
x\to x_\e=x'+\varepsilon(x-x'), &\quad& y\to y_\e=x'+\varepsilon(y-x'), 
\nonumber\\[10pt]
t\to t_\e=\varepsilon^2t, &\quad& \l\to\lambda_\e=\varepsilon^{-2}\lambda,
\nonumber
\eea
and expanding in an asymptotic series in $\varepsilon$.
If one uses a local 
Fourier transform, then the corresponding momenta $\xi\in T^*M$
are large and scale according to 
\bes
\x\to\xi_\e=\varepsilon^{-1}\xi.
\ees
This construction is standard \cite{gilkey95}.
This procedure can be done also in a completely covariant way 
\cite{avra91b,avra98}.

In the case of Laplace type operators, 
the most convenient form of the off-diagonal asymptotics as $t\to 0$, 
among many equivalent forms, is
\cite{avra91b,avra98}
\be
U(t|x,y)\sim (4\pi t)^{-m/2}\exp\left(-{\sigma\over 2t}\right)
\Delta^{1/2}
\sum\limits_{k\ge 0} {(-t)^k\over k!}b_k(x,y),
\label{3.19xxx}
\ee
where $\sigma=\sigma(x,y)=r^2(x,y)/2$ is half the geodesic distance between 
$x$ and $y$, and 
$\Delta=\Delta(x,y)=|g|^{-1/2}(x)|g|^{-1/2}(y)
\det(-\partial^x_\mu\partial^y_{\nu}\sigma(x,y))$ is
the corresponding Van Vleck-Morette determinant. 
The functions $b_k(x,y)$ are called the {\em off-diagonal heat 
coefficients}. These coefficients 
satisfy certain differential recursion relations.  Expanding each
coefficient
in a covariant Taylor series near the 
diagonal, one gets a recursively solvable system of algebraic
equations on the Taylor coefficients \cite{avra91b}.
The diagonal values give the local heat kernel coefficients
$a_k(x)=b_k(x,x)$.
However, in the general case of non-Laplace type operators, 
it is very difficult
to follow this approach, since the 
Ansatz for the off-diagonal heat kernel
asymptotics (\ref{3.19xxx}) does not apply; the correct Ansatz would be much
more complicated.  For this reason,
we employ the approach
of pseudo-differential operators (or, roughly speaking, local
Fourier transforms).  An alternative
approach, which generalizes the 
Ansatz (\ref{3.19xxx}), is developed in Section \ref{ten}.

\section{Gaussian integrals}
\setcounter{equation}0

The {\em Gaussian average} of a function $f(\x)$
on $\RR^m$ is
\bes
<f>\equiv \int\limits_{\RR^m} {d\xi\over \pi^{m/2}}\,e^{-|\xi|^2}f(\xi)\,.
\ees
The Gaussian average of an exponential function gives
the generating function
\bes
I_0(x)=\int\limits_{\RR^m} {d\xi\over\pi^{m/2}}\,\exp\left(
-|\xi|^2+i\xi\cdot x\right)
=\left<\exp\left(i\xi\cdot {x}\right)\right>
=\exp\left(-{|x|^2\over 4}\right).
\ees
{}Expansion of $I_0(x)$ in a power series in $x$ generates the Gaussian
averages of polynomials
\bes
<1>=1, \qquad <\xi_\mu>=0,\qquad <\xi_\mu\xi_\nu>={1\over 2}g_{\mu\nu},
\ees
\bes
<\xi_{\mu_1}\cdots\xi_{\mu_{2n+1}}>=0,
\ees

$$
<\xi_{\mu_1}\cdots\xi_{\mu_{2n}}>
={(2n)!\over 2^{2n}n!} 
g_{(\mu_1\mu_2}\cdots g_{\mu_{2n-1}\mu_{2n})}.
$$
Furthermore, using the relation 
\be
{1\over |\xi|^{2p}}={1\over\Gamma(p)}\int_0^\infty ds\, s^{p-1}e^{-s|\xi|^2},
 \qquad {\rm Re}\, p>0
\label{trick}
\ee
and analytic continuation, one can obtain the more general formulas
\bes
\left<{\xi_{\mu_1}\cdots\xi_{\mu_{2n+1}}\over |\xi|^{2p}}\right>=0,
\ees
\bes
\left<{\xi_{\mu_1}\cdots\xi_{\mu_{2n}}\over |\xi|^{2p}}\right>
={\Gamma(m/2+n-p)\over \Gamma(m/2+n)}{(2n)!\over 2^{2n}n!} 
g_{(\mu_1\mu_2}\cdots g_{\mu_{2n-1}\mu_{2n})}
\ees
for any $p$ with ${\rm Re}\, p<n+m/2$.
This means, in particular, that
\be
\left<{\xi_{\mu_1}\cdots\xi_{\mu_{2n}}\over|\xi|^{2p}}\right>
={\Gamma(m/2+n-p)\over \Gamma(m/2)}
\left<{\xi_{\mu_1}\cdots\xi_{\mu_{2n}}\over
|\xi|^{2n}}\right>,
\label{NonHomGauss}
\ee
\be
\left<{\xi_{\mu_1}\cdots\xi_{\mu_{2n}}\over
|\xi|^{2n}}\right>
={\Gamma(m/2)\over \Gamma(m/2+n)}{(2n)!\over 2^{2n}n!} 
g_{(\mu_1\mu_2}\cdots g_{\mu_{2n-1}\mu_{2n})}\,.
\label{HomGauss}
\ee
Note that if a function $f$ depends only on 
$\xi/|\xi|$, that is, if it is homogeneous of order
$0$, then the average introduced above
is a constant multiple of 
the average over the unit $(m-1)$-sphere $S^{m-1}$. 

We will also need to compute Fourier integrals of the form
\bea
I_{\mu_1\dots\mu_{2n}}(x)&=&
\int_{\RR^m} {d\xi\over\pi^{m/2}} \exp({i\xi\cdot x-|\xi|^2})
{\xi_{\mu_1}\cdots\xi_{\mu_{2n}}\over|\xi|^{2n}}
\nonumber\\[11pt]
&=&\left<{\xi_{\mu_1}\cdots\xi_{\mu_{2n}}\over |\xi|^{2n}}
\exp\left(i\xi\cdot{x}\right)\right>
\eea
The trace of the symmetric $2n$-form $I_{2n}$ over any two indices is
$2(n-1)$-form $I_{2n-2}$:
$$
g^{\mu_{2n-1}\mu_{2n}}I_{\mu_1\cdots\mu_{2n-2}\mu_{2n-1}\mu_{2n}}
=I_{\mu_1\cdots\mu_{2n-2}}.
$$
Therefore, the total trace of the symmetric form $I_{2n}$ is
$I_0$:
$$
\tr_g I_{2n}(x)=I_0(x)=\exp\left(-{|x|^2\over 4}\right)
$$
By rescaling $\xi\to \sqrt t\xi$ one can show that 
this integral satisfies the equation
\bes
t^{m/2}\partial_t^n\left[t^{-m/2}I_{\mu_1\dots\mu_{2n}}
\left({x\over\sqrt{t}}\right)\right]
=\partial_{\mu_1}\cdots\partial_{\mu_{2n}}
\exp\left(-{|x|^2\over 4t}\right).
\ees
Taking into account the obvious asymptotic condition 
$\lim_{t\to\infty}[t^{-m/2}I_n(x/\sqrt{t})]=0$,
by multiple integration we get
\bea
I_{\mu_1\dots\mu_{2n}}(x)&=&
{(-1)^n\over (n-1)!}\int_1^\infty ds\, {(s-1)^{n-1}}{s^{-m/2}}
\partial_{\mu_1}\cdots\partial_{\mu_{2n}}
\exp\left(-{|x|^2\over 4s}\right).\nonumber
\eea
By changing the variable, $s=1/u$, we can rewrite this in the form
\be
I_{\mu_1\dots\mu_{2n}}(x)=
{(-1)^n\over (n-1)!}
\int_0^1 du\, u^{m/2-n-1}(1-u)^{n-1}
\partial_{\mu_1}\cdots\partial_{\mu_{2n}}
\exp\left(-{|x|^2\over 4}\,u\right).
\label{579}
\ee
If $m$ is even this can be computed in elementary functions.
 
In the case $n<m/2$, one can interchange the order of the
integration and differentiation.
Then by using the formula
\bea
&&\int_0^1 du\, u^{a-1}(1-u)^{b-a-1}e^{zu} = 
{\Gamma\left(a\right)\Gamma(b-a)
\over\Gamma(b)}
\ {}_1F_1\left(a; b; z\right),  
\nonumber
\eea
(with ${\rm Re}\, b > {\rm Re}\, a>0$)
where 
$$
{}_1F_1(a;b;z)=\sum\limits_{k=0}^\infty {\Gamma(a+k)\Gamma(b)
\over\Gamma(a)\Gamma(b+k)k!}\,z^k
$$
is the confluent hypergeometric function, we obtain
\bea
I_{\mu_1\dots\mu_{2n}}(x)&=&
(-1)^n\,\partial_{\mu_1}\cdots\partial_{\mu_{2n}}\,
\tilde\Phi_n\left(-{|x|^2\over 4}\right),\nonumber
\eea
where
$$
\tilde\Phi_n(z)={\Gamma(m/2-n)\over\Gamma(m/2)}\,{}_1F_1\left({m\over 2}-n;
{m\over 2};z\right)=
\sum\limits_{k=0}^\infty {\Gamma(m/2-n+k)\over\Gamma(m/2+k)\, k!}\,z^k
$$
(with $n<m/2$).

If $n\ge m/2$, then this formula cannot be applied directly. However,
it is still valid if one does analytic continuation in the dimension
$m$. The physical (integer) value of the dimension should be put
after computing the derivatives.
Alternatively, one can do the differentiation in (\ref{579}) explicitly
before the integration. This is equivalent to subtracting the first $n$ terms
of the Taylor series of the exponential.  This subtraction is clearly harmless,
since the $2n$-th derivatives of these terms vanish. However, this
makes the integral finite, and justifies the interchange of the differentiation and
integration.  The result applies for {\it any} $n$.
In this way one obtains
\bea
I_{\mu_1\dots\mu_{2n}}(x)&=&
(-1)^n\,\partial_{\mu_1}\cdots\partial_{\mu_{2n}}
\Phi_{n}\left(-{|x|^2\over 4}\right),\label{Inxxx}
\eea
where $\Phi_n(z)$ is obtained from the function $\tilde\Phi_n(z)$ by
substracting the first $n$ terms of the power series:
\bea
\Phi_n(z)&=&{\Gamma(m/2-n)\over\Gamma(m/2)}\,{}_1F_1\left({m\over 2}-n;
{m\over 2};z\right)
-\sum\limits_{k=0}^{n-1}{\Gamma(m/2-n+k)\over\Gamma(m/2+k)\,k!}\,z^k
\nonumber\\
&=&\sum\limits_{k=n}^\infty {\Gamma(m/2-n+k)\over\Gamma(m/2+k)\,k!}\,z^k
\label{phin}
\eea

\section{Leading order off-diagonal heat kernel asymptotics}\label{seven}
\setcounter{equation}0

In general, the leading order resolvent and the heat kernel are
determined by the leading symbol of the operator $F$: 
\bes
G_0(\lambda|x,y)=\int {d\xi\over (2\pi)^m}
e^{i\xi\cdot(x-y)}[A(\xi)-\lambda\,\II_{\cal V}]^{-1}, 
\ees 
\bes 
U_0(t|x,y)=\int
{d\xi\over (2\pi)^m} e^{i\xi\cdot(x-y)}\exp[-tA(\xi)], 
\ees 
where
$\xi\in T^*M$, $\xi\cdot (x-y)=\xi_\mu (x^\mu-y^\mu)$, and $d\xi$ is
Lebesgue measure on $\RR^m$. Here and everywhere below all integrals over 
$\xi$ below will be over the whole $\RR^m$.

Writing the leading symbol in terms of the projections from (\ref{AtoPi}),
it is not difficult to obtain
\bes
G_0(\lambda|x,y)=\sum_{i=1}^s
\int {d\xi\over (2\pi)^m} 
e^{i\xi\cdot(x-y)}{\Pi_i(\xi)\over \mu_i|\xi|^2-\lambda},
\ees
\bes
U_0(t|x,y)=\sum_{i=1}^s
\int {d\xi\over (2\pi)^m} e^{i\xi\cdot(x-y)-t\mu_i|\xi|^2}\Pi_i(\xi).
\ees

The problem is now to compute these integrals.
If we are only interested in traces, then by 
(\ref{Qxxx}), we can easily compute
\bes
\tr_{\cal V} U_0(t|x,y)=\sum_{i=1}^s d_i (4\pi t\mu_i)^{-m/2}
\exp\left(-{|x-y|^2\over 4t\mu_i}\right).
\ees
That is, the trace of the leading order term in the heat 
kernel asymptotics is a weighted linear combination of the
scalar leading order heat kernel term with scaled times, $t\to \mu_i t$.  
Similarly, we can relate the trace of the resolvent at leading order
to resolvents of Laplace type operators:
\bes
\tr_{\cal V}G_0(\lambda|x,y)=
\sum_{i=1}^s d_i (2\pi \mu_i)^{-m/2}
\left({-\lambda\mu_i \over |x-y|^2}\right)^{(m-2)/4}
K_{(m-2)/2}\left(\sqrt{{-\lambda\over \mu_i}}|x-y|\right),
\ees
where $K_p(z)$ is the modified Bessel function.

Applying the methods described in the previous subsection we can compute
the heat kernel and resolvent {\it before taking the trace}:
\bes
U_0(t|x,y)=\sum_{i=1}^s 
(4\pi t \mu_i)^{-m/2}
\left<\exp\left[i\xi\cdot{(x-y)\over \sqrt{t\mu_i}}\right]\Pi_i(\xi)\right>.
\ees
Using the decomposition of the projections into spherical 
harmonics (\ref{321}), we obtain
\bea
U_0(t|x,y)&=&\sum_{1\le i,k \le s}
(4\pi t \mu_i)^{-m/2}c_{ik}
a^{(\mu_1\mu_2}\cdots a^{\mu_{2k-3}\mu_{2k-2})}
(-t\mu_i)^{k-1}
\nonumber\\[12pt]
&\times&\partial_{\mu_1}\cdots\partial_{\mu_{2k-2}}
\Phi_{k-1}\left(-{|x-y|^2\over 4t\mu_i}\right),
\nonumber\\[12pt]
&=&
\sum_{1\le i,k \le s}
(4\pi t \mu_i)^{-m/2}c_{ik}
(t\mu_i F_0)^{k-1}\Phi_{k-1}\left(-{|x-y|^2\over 4t\mu_i}\right),
\eea
\bea
G_0(\lambda|x,y)&=&
\sum_{1\le i,k \le s}
(4\pi \mu_i)^{-m/2}c_{ik}
(\mu_i F_0)^{k-1}
\nonumber\\
&\times&
\int_0^\infty dt\, e^{t\lambda}t^{-m/2+k-1}
\Phi_{k-1}\left(-{|x-y|^2\over 4t\mu_i}\right)\;,
\nonumber
\eea
where $F_0=a^{\mu\nu}\partial_\mu\partial_\nu$
and the functions $\Phi_n(z)$ are defined in previous section; they
are given by (\ref{phin}).
We would like to stress that these formulas can be presented {\it locally}
in a ``covariantized'' form. This is effectively achieved by replacing
$|x-y|^2$ by $2\sigma(x,y)$ and $(x-y)^\mu$ by $\nabla^\mu\sigma(x,y)$
and adding a factor $\Delta^{1/2}(x,y)$ (for details, see \cite{avra91b});
the objects $\sigma$ and $\Delta$ being defined after equation (\ref{3.19xxx}).

\section{The heat kernel coefficient $a_0$}
\setcounter{equation}0

The leading order term of the diagonal heat kernel asymptotics
can now be written as
\bes
U^{\rm diag}_0(t):=U_0(t|x,x)=(4\pi t)^{-m/2}\sum_{i=1}^s \mu_i^{-m/2}
\left<\Pi_i\right>.
\ees
This gives the coefficient $a_0\,$:
\be
a_0=\sum_{i=1}^s \mu_i^{-m/2}
\left<\Pi_i\right>.
\label{a0xxx}
\ee
The Gaussian average is computed by using (\ref{HomGauss}):
\bea
<\Pi_i>=\Pi_{i(0)}&=&
\sum_{k=1}^s c_{ik}{\Gamma(m/2) (2k-2)!\over \Gamma(m/2+k-1) 2^{2k-2}(k-1)!}
{\rm tr}_g \vee^{k-1} a\,,
\label{PiGauss}
\eea
where $\vee^n$ is the symmetrized tensor power of symmetric form, ${\rm tr}_g$
 is the total trace of a symmetric form defined in (\ref{tottr})
and the constants $c_{ik}$ are defined by (\ref{cik}).

{}From this, we have the formula
\bes
\tr_\cV a_0=\sum_{i=1}^s {d_i\over \mu_i^{m/2}}
\ees
for the trace of the heat coefficient.

Note that if our bundle $\cV$ is associated to an 
{\it irreducible} representation of $\spin(m)$,
the averages of the projections, being $\spin(m)$-invariant endomorphisms,
are 
proportional to the identity, by Schur's Lemma.
The exact proportionality constant may be determined by taking the trace.
We have:
\bes
<\Pi_{i}>={d_i\over d}\II_{\cal V}, 
\qquad a_0={1\over d}\sum_{i=1}^s {d_i\over \mu_i^{m/2}}\II_{\cal V}.
\ees

These formulas point up a new 
feature of non-Laplace type operators; one which complicates life somewhat.
Whereas the dimension 
dependence 
of the heat coefficients of Laplace type operators is isolated in the
overall factor of $(4\pi)^{-m/2}$, the dimension dependence for 
NLT operators is more complicated.

\section{The heat kernel coefficient $A_1$}\label{nine}
\setcounter{equation}0

As we have seen, even the computation  of the leading order heat kernel
requires significant effort in the case of non-Laplace principal
part.  This indicates that 
the calculation of higher-order
coefficients will be a challenging task. 
In this paper we will compute the coefficient
$A_1$. 
Since $A_1$ is the integral of $\tr_\cV a_1$ by 
(\ref{Ak}), it suffices to compute the local
coefficient $a_1$ modulo trace-free endomorphisms. 
By elementary invariant theory, $a_1$ has the form
\be
a_1=\sum_{k\ge 0} Q_{(k)\,\mu_1\cdots\mu_k}q P_{(k)}^{\mu_1\cdots\mu_k}
+H_1R
+H_2^{\mu\nu} R_{\mu\nu}
+H_3^{\mu\nu\alpha\beta}
R_{\mu\nu\alpha\beta}
\label{987xxx}
\ee
where $Q_{(k)}$, $P_{(k)}$, and $H_i$ are $\End({\cal V})$-valued tensors, 
and $R$, $R_{\mu\nu}$, and
$R_{\mu\nu\alpha\beta}$ are the scalar, Ricci, and Riemann 
curvatures.

For the trace we have a similar formula,
\bes
\tr_{\cal V}a_1=\tr_{\cal V}(h_0q)+h_1R
+h_2^{\mu\nu} R_{\mu\nu}
+h_3^{\mu\nu\alpha\beta}
R_{\mu\nu\alpha\beta}
\ees
where $h_i\equiv \tr_{\cal V} H_i$ ($i=1,2,3$) are some tensors. 
By invariant theory, we may
conclude that these tensors have the following form:
\bes
h_0=\sum_{k\ge 0} P_{(k)}^{\mu_1\cdots\mu_k}Q_{(k)\,\mu_1\cdots\mu_k},
\ees
\bes
h_1=c_1,\qquad
h_2^{\mu\nu}=c_2 g^{\mu\nu}, \qquad
h_3^{\mu\nu\alpha\beta}=c_3g^{\mu\alpha}g^{\nu\beta}
+c_4g^{\mu\beta}g^{\nu\alpha}
+c_5g^{\mu\nu}g^{\alpha\beta},
\ees
where $c_i$ are some constants.

Note that for irreducible representations
the endomorphisms $Q_0$, $P_0$, 
$h_0$ and $H_1$ are proportional to the identity endomorphism;
in particular, $h_0=c_0\II_{\cal V}$, $H_1=(c_1/d)\II_{\cal V}$.

Taking this into account, we obtain  
\bes
\tr_{\cal V}a_1=\tr_{\cal V}(h_0q)+\beta R,
\ees
where
\bes
\beta=c_1+c_2+c_3-c_4.
\ees
Thus it suffices to compute the constant $\beta$ and the
endomorphism $h_0$.

To compute $h_0$, we can employ a by now standard variational principle.
If the operator $F(\varepsilon)$ depends (in a suitably estimable way)
on a parameter $\varepsilon$,
then 
\bes
{\partial\over\partial \varepsilon}\Tr_{L^2}\exp(-tF)
=-t\Tr_{L^2}(\partial_\varepsilon F)\exp(-tF).
\ees
If our variation comes from scaling $q$,
$$
q\to \varepsilon q,
$$ 
then
$(\partial_\varepsilon F)=q$.
Expanding both sides in powers of $t$ and comparing coefficients of like
powers, we obtain
\bes
\partial_\varepsilon A_1=\int_Md\vol(x)\tr_{\cal V}(a_0q),
\ees
so that
$$
A_1=\int_Md\vol(x)\left[\tr_{\cal V}(a_0q)+\beta R\right]\;,
$$
and, therefore, $h_0=a_0$,
where $a_0$ is given by (\ref{a0xxx}) and (\ref{PiGauss}).

The computation of the constant $\b$ is considerably more difficult.
It is clear the $\b$ is {\em universal} to our class of operators; thus
we may compute it for any particular operator and manifold, or class of
such.  Accordingly, note that
$\beta$ is equal to the coefficient $a_1$ when $q=0$ and $R=1$:
\bes
\beta=\tr_{\cal V}a_1\Big|_{q=0, R=1}\,.
\ees

The following considerations will be completely local. 
Fix a point $x'$, and compute in normal coordinates centered at $x'$,
with $g_{\mu\nu}(0)=\delta_{\mu\nu}$.
Furthermore, impose by gauge transformation
the {\em Fock-Schwinger} gauge for the connection 1-form ${\cal A}$.
We then have
\bes
[g_{\mu\nu}(x)-\delta_{\mu\nu}]x^\nu=0, \qquad
{\cal A}_{\mu}(x)x^\mu=0.
\ees
Further, we
can expand all quantities in Taylor series about $x'$
and restrict our attention to terms linear in the curvature. 
This gives
\bes
g_{\mu\nu}(x)=\delta_{\mu\nu}-{1\over 3}R_{\mu\alpha\nu\beta}x^\alpha x^\beta
+\dots,
\ees
\bes
\det\,g_{\mu\nu}(x)=1-{1\over 3}R_{\alpha\beta}x^\alpha x^\beta
+\dots,
\ees
\bes
{\cal A}_{\mu}(x)=-{1\over 2}{\cal R}_{\mu\alpha}x^\alpha
+\dots.
\ees
Here and below, 
the dots denote higher-order terms in the curvature.
${\cal R}_{\mu\nu}$ is, of course, 
the curvature of the bundle ${\cal V}$, given by
$
{\cal R}_{\mu\nu}=R^{\alpha}{}_{\beta\mu\nu}T^\beta{}_\alpha.
$
The Taylor expansion of the leading symbol section
$a^{\mu\nu}$ is determined by the equation
\bes
\nabla_\mu a^{\alpha\beta}=0.
\ees
{}From this, we get
\bes
a^{\mu\nu}(x)=a^{\mu\nu}
+{1\over 3}a^{\lambda(\mu}R^{\nu)}{}_{\alpha\lambda\beta}x^\alpha x^\beta
+\dots.
\ees
Here and below, we denote $a^{\mu\nu}(0)$ 
simply by $a^{\mu\nu}$.
By the above, the potential term $q$ will only
enter the calculation through $q(0)$, which we denote simply
by $q$:
\bes
q(x)=q+\dots.
\ees

Since the constant $\b$ is universal, we are free to
compute in the case of
a constant curvature metric, i.e.
\be
R_{\mu\nu\alpha\beta}={R\over m(m-1)}(g_{\mu\alpha}g_{\beta\nu}
-g_{\mu\beta}g_{\alpha\nu}),
\label{9105xxx}
\ee
\bes
R_{\mu\nu}={R\over m}g_{\mu\nu}, \qquad R=\const.
\ees

Now let us take the {\em total} symbol
of our operator in normal coordinates,
$\sigma(F|x,\xi)$, and expand in 
a Taylor series:
\bes
\sigma(F|x,\xi)=\sigma_L(F|0,\xi)
+\sigma(F_1|x,\xi)+\dots,
\ees
where
\bes
\sigma_L(F|0,\xi)=a^{\mu\nu}\xi_\mu\xi_\nu\equiv A(\xi),
\ees
\bes
\sigma(F_1|x,\xi)=-X^{\mu\nu}{}_{\alpha\beta}x^\alpha x^\beta\xi_\mu\xi_\nu
+iY^\mu{}_\alpha x^\alpha\xi_\mu
+q,
\ees
\bes
X^{\mu\nu}{}_{\alpha\beta}
=-{1\over 3}a^{\lambda(\mu}R^{\nu)}{}_{(\alpha|\lambda|\beta)},
\ees
\bes
Y^\mu{}_\alpha={2\over 3}a^{\mu\lambda}R_{\lambda\alpha}
-{1\over 2}[T^{\sigma}{}_{\rho}, a^{\mu\nu}]_+R^\rho{}_{\sigma\alpha\nu},
\ees
and $[A , B]_+=AB+BA$ denotes the anticommutator.

There are many equivalent ways of constructing the heat kernel 
asympotics on the diagonal locally.
Using the Volterra series
\bes
\exp(-tF)=\exp(-tF_0)
-t\int_0^1 d\tau \exp[-t(1-\tau)F_0]F_1\exp[-t\tau F_0]+\dots.
\ees
and the formula for the heat kernel on the diagonal
\bes
U^{\rm diag}(t)
=\int{d\xi\,\over (2\pi)^{m}} e^{-i\xi\cdot x}
\exp(-tF)e^{i\xi\cdot x}\Big|_{x=0},
\ees
we get
\bes
U^{\rm diag}(t)
=\int{d\xi\,\over (2\pi)^{m}}
\left\{e^{-tA(\xi)}
-t\int_0^1 d\tau e^{-t(1-\tau)A(\xi)}
\hat F_1e^{-t\tau A(\xi)}
+\dots\right\},
\ees
where
\bes
\hat F_1=\sigma(F_1|i\partial_\xi,\xi)
=X^{\mu\nu}{}_{\alpha\beta}\partial_\xi^\alpha 
\partial_\xi^\beta\xi_\mu\xi_\nu
-Y^\mu{}_\alpha \partial_\xi^\alpha\xi_\mu
+q.
\ees

Finally, by scaling the integration variable $\xi\to t^{-1/2}\xi$, 
we obtain the standard
asymptotic expansion of the heat kernel on the 
diagonal  (\ref{566xxx}), with the coefficients
\bes
a_0=\int {d\xi\over \pi^{m/2}}\, e^{-A(\xi)},
\ees
\bes
a_1=\int {d\xi\over \pi^{m/2}}\,
\int_0^1 d\tau e^{-(1-\tau)A(\xi)}\hat F_1e^{-\tau A(\xi)}.
\ees
In particular, we recover the formula for $a_0$ derived in 
the previous section.

To integrate by parts in 
this formula, we need to know how to differentiate the 
exponential $e^A$. This can be
done via the Duhamel formula
\bes
\partial_\xi^\alpha e^{-\tau A(\xi)}
=-2\int_0^\tau ds\, e^{-(\tau-s)A(\xi)}J^\alpha(\xi) e^{-sA(\xi)},
\ees
where
\bes
J^\alpha(\xi)=a^{\alpha\beta}\xi_\beta.
\ees
The contraction of this with $\x$ leads to a much simpler formula:
\bes
\xi_\alpha\partial_\xi^\alpha e^{-\tau A(\xi)}
=-2\tau A(\xi) e^{-\tau A(\xi)}.
\ees

The exponential itself is computed by using the projections $\P_i\,$:
\bes
e^{-\tau A(\xi)}=\sum_{i=1}^s e^{-\tau\mu_i|\xi|^2}\Pi_i(\xi).
\ees
Note that when computing $\xi$-derivatives one can,
using integration by parts, act in either direction.
This trick
can be used to avoid having 
to compute the second derivative of the exponential $e^A$. 
First, we rewrite $\hat F_1$ in the form
\bes
\hat F_1=\partial_\xi^\alpha X_{\alpha\beta}\partial_\xi^\beta
+\partial_\xi^\alpha L_\alpha
+L_\alpha\partial_\xi^\alpha
+Q,
\ees
where
\bes
X_{\alpha\beta}(\xi)=X^{\mu\nu}{}_{\alpha\beta}\xi_\mu\xi_\nu
=-{1\over 3}a^{\lambda(\mu}R^{\nu)}{}_{(\alpha|\lambda|\beta)}\xi_\mu\xi_\nu,
\ees
\bes
L_\alpha(\xi)=L^{\mu}{}_{\alpha}\xi_\mu,
\ees
\bes
L^\mu{}_\alpha=-{1\over 4}a^{\mu\lambda}R_{\lambda\alpha}
+{1\over 12}a^{\lambda\nu}R^{\mu}{}_{\lambda\alpha\nu}
+{1\over 4}R^\rho{}_{\sigma\alpha\nu}[T^{\sigma}{}_{\rho}, a^{\mu\nu}]_+.
\ees
\bes
Q=q-{1\over 6}a^{\mu\nu}R_{\mu\nu}.
\ees
Now, integrating by parts over $\xi$ (and changing $\tau$ to 
$1-\tau$ in some places)
we get
\bea
a_1&=&\int {d\xi\over \pi^{m/2}}\,\,\Biggl\{
-4\int_0^1d\tau\int_0^{1-\tau} ds_1 
\int_0^{\tau} ds_2\,e^{-(1-\tau-s_1)A}
\nonumber\\
&&\times
J^\alpha e^{-s_1 A}X_{\alpha\beta}
e^{-(\tau-s_2)A}J^\beta e^{-s_2A}
\nonumber\\
&&
+2\int_0^1 d\tau\int_0^{\tau} ds\,
\Bigl[e^{-(\tau-s)A}J^\alpha e^{-sA}L_\alpha e^{-(1-\tau) A}
\nonumber\\
&&
-e^{-(1-\tau)A}L_\alpha e^{-(\tau-s)A}J^\alpha e^{-s A}
\Bigr]
\nonumber\\
&&
+\int_0^1 d\tau\,e^{-(1-\tau)A}Qe^{-\tau A}
\Biggr\}.\label{trva1}
\eea

By separating the curvature factors in (\ref{trva1}), 
one can compute each of the
tensors $H_i$ entering (\ref{987xxx}).
Note that for a Laplace type operator (the case 
$a^{\mu\nu}=g^{\mu\nu}\II$) we have
$A=|\xi|^2$, $J^\mu=\xi^\mu$, 
$X_{\alpha\beta}=-(1/3)R^\mu{}_\alpha{}^\nu{}_\beta\xi_\mu\xi_\nu$, 
$L_\alpha=-(1/6)R^\mu{}_\alpha\xi_\mu$, $Q=q-(1/6)R\II$.
Therefore the first two terms (with $X$ and $L$)
vanish in the Laplace type case, and we get the well known result
\bes
a_1=q-{1\over 6}R\II_{\cal V}.
\ees

Note that the tensors $H_i$ depend {\it only}
on the leading symbol, i.e. on $a^{\mu\nu}$. 
In principle, it is possible to compute them explicitly
by using the representation of the $e^A$ in terms of projections 
and Gaussian averages.

We shall not do this explicitly, but rather compute only the {\it trace}
of $a_1$.  The number of projections 
involved in this calculation is less by one.
Computing in the case of constant curvature (\ref{9105xxx}),
and attaching a tilde to $X$, $L$ and $Q$ 
in this case, we have
\bes
\tilde X_{\alpha\beta}(\xi)
=-{R\over 3m(m-1)}\left[Ag_{\alpha\beta}-J_{(\alpha}\xi_{\beta)}\right],
\ees
\be
\tilde L_\alpha(\xi)=-{R\over 12m(m-1)}\left\{\bar a \xi_\alpha-(3m-2)J_\alpha
+6[T_{[\nu\alpha]}, J^\nu]_+\right\},
\label{9130xxx}
\ee
\bes
\tilde Q=q-{R\over 6m}\bar a,
\ees
where
\bes
\bar a=g_{\mu\nu}a^{\mu\nu}.
\ees
Now taking the trace and changing $s$ to $\tau-s$ in the second integral, 
we obtain
\bea
\tr_{\cal V}a_1&=&\tr_{\cal V}\int {d\xi\over \pi^{m/2}}\,\,\Biggl\{
-4\int_0^1d\tau\int_0^{1-\tau} ds_1 \int_0^{\tau} ds_2\,
e^{-(\tau-s_2)A}
\nonumber\\
&&\times
J^\beta e^{-(1-\tau-s_1+s_2)A}J^\alpha e^{-s_1 A}
\tilde X_{\alpha\beta}
\nonumber\\
&&
+2\int_0^1 d\tau\int_0^{\tau} ds\,
\Bigl[e^{-(1-s)A}J^\alpha e^{-sA}
-e^{-sA}J^\alpha e^{-(1-s)A}
\Bigr]\tilde L_\alpha
\nonumber\\
&&
+e^{-A}\tilde Q
\Biggr\}.
\eea
In the second multiple, the $s$-integration may be 
accomplished explicitly, to give
\bea
\tr_{\cal V}a_1&=&\tr_{\cal V}\int {d\xi\over \pi^{m/2}}\,\,\Biggl\{
-4\int_0^1d\tau\int_0^{1-\tau} ds_1 \int_0^{\tau} ds_2\,
e^{-(\tau-s_2)A}
\nonumber\\
&&\times
J^\beta e^{-(1-\tau-s_1+s_2)A}J^\alpha e^{-s_1 A}
\tilde X_{\alpha\beta}
\nonumber\\
&&
+2\int_0^1 d\tau\,\tau
\Bigl[e^{-\tau A}J^\alpha e^{-(1-\tau) A}
-e^{-(1-\tau)A}J^\alpha e^{-\tau A}
\Bigr]\tilde L_\alpha
\nonumber\\
&&
+e^{-A}\tilde Q
\Biggr\}.
\eea

Now let us consider the different contributions separately.

\medskip
\leftline{\bf $Q$ contribution.}
\medskip\noindent
The $Q$ contribution has the form
\bes
\tr_{\cal V}\int {d\xi\over \pi^{m/2}}\,\,e^{-A}\tilde Q
=\tr_{\cal V}\sum_{i=1}^s 
\int {d\xi\over \pi^{m/2}}\,e^{-\mu_i|\xi|^2}\Pi_i\tilde Q
\ees
By changing the integration variable $\xi$ to $(\mu_i)^{-1/2}\xi$, we obtain
a Gaussian average:
\bes
\tr_{\cal V}\sum_{i=1}^s \mu^{-m/2}_i<\Pi_i>
\left(q-{1\over 6m}\bar a R\right)
=\tr_{\cal V} a_0\left(q-{1\over 6m}\bar a R\right).
\ees

\medskip
\leftline{\bf $L$ contribution.}
\medskip\noindent
This contribution has the form
\bea
2\tr_{\cal V}\int {d\xi\over \pi^{m/2}}\,
\int_0^1 d\tau\,\tau
\Bigl[e^{-\tau A}J^\alpha e^{-(1-\tau) A}
-e^{-(1-\tau)A}J^\alpha e^{-\tau A}
\Bigr]\tilde L_\alpha\,.
\eea
Using (\ref{9130xxx}), we get
\bea
&&-{R\over 6m(m-1)}\tr_{\cal V}\int {d\xi\over \pi^{m/2}}\,
\int_0^1 d\tau\,\tau
\Bigl[e^{-\tau A}J^\alpha e^{-(1-\tau) A}
-e^{-(1-\tau)A}J^\alpha e^{-\tau A}
\Bigr]
\nonumber\\[11pt]
&&
\times\left(\bar a \xi_\alpha-(3m-2)J_\alpha
+6[T_{[\nu\alpha]}, J^\nu]_+\right).
\eea
Noting that $J^\alpha\xi_\alpha=A$, we see that the term in $L_\alpha$
proportional to $\xi_\alpha$ does not contribute.
The term proportional to $J_\alpha$ gives
\bes
R{(3m-2)\over 6m(m-1)}\tr_{\cal V}\int {d\xi\over \pi^{m/2}}\,
\int_0^1 d\tau\,\tau
\Bigl[e^{-\tau A}J^\alpha e^{-(1-\tau) A}
-e^{-(1-\tau)A}J^\alpha e^{-\tau A}
\Bigr]J_\alpha\,.
\ees
This is computed by again introducing the projections $\Pi_i\,$:
\bea
&&R{(3m-2)\over 6m(m-1)}
\sum_{1\le i,k\le s}\int {d\xi\over \pi^{m/2}}\,
\int_0^1 d\tau\,\tau
\Big[e^{-[\mu_i\tau+\mu_k(1-\tau)]|\xi|^2}
\nonumber\\[12pt]
&&
-e^{-[\mu_k\tau+\mu_i(1-\tau)]|\xi|^2}
\Big]
\tr_{\cal V}\Pi_iJ^\alpha \Pi_kJ_\alpha\,.
\nonumber
\eea
By changing the integration variable $\xi$ to 
$[\mu_i\tau+\mu_k(1-\tau)]^{-1/2}\xi$, we may 
express this in terms of a Gaussian average:
\bes
R{(3m-2)\over 6m(m-1)}
\sum_{1\le i,k\le s}\kappa_{ik}
\tr_{\cal V}\left<\Pi_iJ^\alpha \Pi_kJ_\alpha\right>,
\ees
where
\bes
\kappa_{ik}=
\int_0^1 d\tau\,\tau\left\{[\mu_i\tau+\mu_k(1-\tau)]^{-(m+2)/2}
-[\mu_k\tau+\mu_i(1-\tau)]^{-(m+2)/2}
\right\}.
\ees
Clearly $\kappa_{ik}=-\kappa_{ki}\,$; hence, for $i=k$
\bes
\kappa_{ii}=0.
\ees
For $i\ne k$ we compute
\be
\kappa_{ik}=-{\Gamma(m/2-1)\over \Gamma(m/2+1)(\mu_i-\mu_k)}
\left\{{m\over 2}\left(\mu_i^{-m/2}+\mu_k^{-m/2}\right)
+{\mu_i^{-m/2+1}-\mu_k^{-m/2+1}\over \mu_i-\mu_k}
\right\}.
\label{9144xxx}
\ee

Similarly, the contribution of the term with $T$ is computed to be
\bea
&&-{R\over m(m-1)}\tr_{\cal V}\int {d\xi\over \pi^{m/2}}\,
\int_0^1 d\tau\,\tau
\Bigl[e^{-\tau A}J^\alpha e^{-(1-\tau) A}
\nonumber\\[12pt]
&&
-e^{-(1-\tau)A}J^\alpha e^{-\tau A}
\Bigr]
[T_{[\nu\alpha]}, J^\nu]_+
\nonumber\\[11pt]
&&=-{R\over m(m-1)}
\tr_{\cal V}\sum_{1\le i,k\le s}\kappa_{ik}
\left<\Pi_iJ^\alpha \Pi_k [T_{[\nu\alpha]}, J^\nu]_+
\right>
\nonumber\\[11pt]
&&=-{R\over m(m-1)}
\tr_{\cal V}\sum_{1\le i,k\le s}\kappa_{ik}
\left(\left<J^\nu\Pi_iJ^\alpha \Pi_k\right>
+\left<\Pi_iJ^\alpha \Pi_k J^\nu\right>\right)T_{[\nu\alpha]}.
\nonumber
\eea
Thus the total contribution of $L$ is
\bea
&&
R{1\over 6m(m-1)}
\tr_{\cal V}\sum_{1\le i,k\le s}\kappa_{ik}
\Bigg\{
(3m-2)\left<\Pi_iJ^\alpha \Pi_kJ_\alpha\right>
\nonumber\\[11pt]
&&
-6\left(\left<J^\nu\Pi_iJ^\alpha \Pi_k\right>
+\left<\Pi_iJ^\alpha \Pi_k J^\nu\right>\right)T_{[\nu\alpha]}
\Bigg\}.
\nonumber
\eea

\medskip
\leftline{\bf $X$ contribution.}
\medskip\noindent
This term has the form
\bea
&&R{4\over 3m(m-1)}\tr_{\cal V}\int {d\xi\over \pi^{m/2}}\,\,\Biggl\{
\int_0^1d\tau\int_0^{1-\tau} ds_1 \int_0^{\tau} ds_2\,
e^{-(\tau-s_2)A}
\nonumber\\[11pt]
&&\times
J^\beta e^{-(1-\tau-s_1+s_2)A}J^\alpha e^{-s_1 A}
\left[Ag_{\alpha\beta}-J_{(\alpha}\xi_{\beta)}\right].
\nonumber
\eea
We consider first the term proportional to $A$. 
Because $A$ commutes with the exponent,
we have
\bea
&&
R{4\over 3m(m-1)}\tr_{\cal V}\int {d\xi\over \pi^{m/2}}\,
\int_0^1d\tau\int_0^{1-\tau} ds_1 \int_0^{\tau} ds_2\,
Ae^{-(\tau-s_2+s_1)A}
\nonumber\\[11pt]
&&\times
J^\alpha e^{-(1-\tau-s_1+s_2)A}J_\alpha\,.
\eea
Introducing the projections again, we get
\bea
&&R{4\over 3m(m-1)}\tr_{\cal V}
\sum_{1\le i,k\le s}
\int {d\xi\over \pi^{m/2}}\,
\int_0^1d\tau\int_0^{1-\tau} ds_1 \int_0^{\tau} ds_2\,
\nonumber\\[11pt]
&&\times
e^{-[(\tau-s_2+s_1)\mu_i +(1-\tau-s_1+s_2)\mu_k]|\xi|^2}
\mu_i|\xi|^2
\Pi_iJ^\alpha \Pi_k J_\alpha\,.\nonumber
\eea
Scaling, $\xi\to [(\tau-s_2+s_1)\mu_i +(1-\tau-s_1+s_2)\mu_k]^{-1/2}\xi$,
we obtain
\bea
&&R{4\over 3m(m-1)}\tr_{\cal V}
\sum_{1\le i,k\le s}
\rho_{ik}
\mu_i\left<|\xi|^2\Pi_iJ^\alpha \Pi_k J_\alpha\right>,
\nonumber
\eea
where
\bes
\rho_{ik}
=\int_0^1d\tau\int_0^{1-\tau} ds_1 \int_0^{\tau} ds_2\,
 [(\tau-s_2+s_1)\mu_i +(1-\tau-s_1+s_2)\mu_k]^{-(m+4)/2}.
\ees
By changing $\tau$ to $1-\tau$ and switching the roles of 
$s_1$ and $s_2\,$, we see that
\bes
\rho_{ik}=\rho_{ki}.
\ees
{}For $i=k$, we easily obtain
\bes
\rho_{ii}
={1\over 6} \mu_i^{-m/2-2}\,.
\ees
{}For $i\ne k$ we compute that
\bea
\rho_{ik}&=&{\Gamma(m/2-1)\over \Gamma(m/2+2)(\mu_i-\mu_k)^2}
\Bigg\{(m-4)\left(\mu_i^{-m/2}+\mu_k^{-m/2}\right)
\nonumber\\[11pt]
&&
+2\,{\mu_i^{-m/2+1}-\mu_k^{-m/2+1}\over \mu_i-\mu_k}
\Bigg\}.
\eea
Furthermore, the term proportional to $\xi$ in the $X$ contribution is
\bea
&&-R{4\over 3m(m-1)}\tr_{\cal V}\int {d\xi\over \pi^{m/2}}\,
\int_0^1d\tau\int_0^{1-\tau} ds_1 \int_0^{\tau} ds_2\,
\nonumber\\[11pt]
&&
e^{-(\tau-s_2)A}J^\beta e^{-(1-\tau-s_1+s_2)A}J^\alpha e^{-s_1 A}
J_{(\alpha}\xi_{\beta)},\nonumber
\eea
Remembering that $J^\alpha\xi_\alpha=A$ and changing the variables
of integration, we have
\bea
&&-R{4\over 3m(m-1)}\tr_{\cal V}\int {d\xi\over \pi^{m/2}}\,
\int_0^1d\tau {1\over 2}\tau^2
Ae^{-\tau A}J^\alpha e^{-(1-\tau) A}
J_{\alpha}\,.\nonumber
\eea
Introducing the projections $\Pi_i\,$,
this becomes
\bes
-R{4\over 3m(m-1)}\tr_{\cal V}
\sum_{1\le i,k\le s}\int {d\xi\over \pi^{m/2}}\,
\int_0^1d\tau {1\over 2}\tau^2
\mu_i e^{-[\tau\mu_i+(1-\tau)\mu_k]|\xi|^2}
|\xi|^2\Pi_iJ^\alpha \Pi_kJ_{\alpha}\,.
\ees
Finally, scaling $\xi\to [\tau\mu_i+(1-\tau)\mu_k]^{-1/2}\xi$, we get
\bea
&&-R{4\over 3m(m-1)}\tr_{\cal V}
\sum_{1\le i,k\le s}
\gamma_{ik}\mu_i
\left<|\xi|^2\Pi_iJ^\alpha \Pi_kJ_{\alpha}\right>,\nonumber
\eea
where
\bes
\gamma_{ik}=\int_0^1d\tau {1\over 2}\tau^2
[\tau\mu_i+(1-\tau)\mu_k]^{-m/2-2}.
\ees
{}For $i=k$,
\bes
\gamma_{ii}={1\over 6}\mu_i^{-m/2-2}.
\ees
{}For $i\ne k$, we compute that
\bea
\gamma_{ik}
&=&{\Gamma(m/2-1)\over 8\Gamma(m/2+2)(\mu_i-\mu_k)}
\Biggl\{
-{m(m-2)}\mu_i^{-m/2-1}
-{4(m-2)}{\mu_i^{-m/2}\over \mu_i-\mu_k}
\nonumber\\[11pt]
&&
-8\,{\mu_i^{-m/2+1}-\mu_k^{-m/2+1}\over (\mu_i-\mu_k)^2}
\Biggr\}.\nonumber
\eea

Thus, the total $X$ contribution is
\bea
&&R{4\over 3m(m-1)}\tr_{\cal V}
\sum_{1\le i,k\le s}
\mu_i\sigma_{ik}
\left<|\xi|^2\Pi_iJ^\alpha \Pi_kJ_{\alpha}\right>,
\nonumber
\eea
where
\bes
\sigma_{ik}=\rho_{ik}-\gamma_{ik}\,.
\ees
Clearly
\bes
\sigma_{ii}=0.
\ees
For $i\ne k$,
\bea
\sigma_{ik}
&=&{\Gamma(m/2-1)\over 8\Gamma(m/2+2)} 
\Biggl\{
24\,{\mu_i^{-m/2+1}-\mu_k^{-m/2+1}\over (\mu_i-\mu_k)^3}
+8(m-4){\mu_i^{-m/2}+\mu_k^{-m/2}\over  (\mu_i-\mu_k)^2 }
\nonumber\\[11pt]
&&
+{4(m-2)}{\mu_i^{-m/2}\over (\mu_i-\mu_k)^2}
+{m(m-2)}{\mu_i^{-m/2-1}\over  (\mu_i-\mu_k)}
\Biggr\}.
\label{9165xxx}
\eea

\medskip
\leftline{\bf Result.}
\medskip\noindent
Summing up the various contributions,
we get the main result:
\bes
\tr_{\cal V} a_1=\sum_{i=1}^s \mu^{-m/2}_i\tr_{\cal V}<\Pi_i>q+\beta R,
\ees
where
\bea
\beta&=&
-{1\over 6m}\sum_{i=1}^s \mu^{-m/2}_i\tr_{\cal V}<\Pi_i>\bar a 
\nonumber\\[11pt]
&&
+{1\over 6m(m-1)}
\sum_{1\le i,k\le s;\,i\ne k}\kappa_{ik}
\Bigl\{
(3m-2)\tr_{\cal V}\left<\Pi_iJ^\alpha \Pi_kJ_\alpha\right>
\nonumber\\[11pt]
&&
-6\tr_{\cal V}\left[\left<J^\nu\Pi_iJ^\alpha \Pi_k\right>
+\left<\Pi_iJ^\alpha \Pi_k J^\nu\right>\right]T_{[\nu\alpha]}
\Bigr\}
\nonumber\\[11pt]
&&+{4\over 3m(m-1)}
\sum_{1\le i,k\le s;\,i\ne k}
\mu_i\sigma_{ik}
\tr_{\cal V}\left<|\xi|^2\Pi_iJ^\alpha \Pi_kJ_{\alpha}\right>,
\nonumber
\eea
with the constants $\kappa_{ik}$ and $\sigma_{ik}$ given by (\ref{9144xxx})
and (\ref{9165xxx}).

Next we need to compute the Gaussian averages. This can be easily done
by using the formulas (\ref{NonHomGauss}).
First note that the radial part can be always separated by
\bes
\left<|\xi|^{2p}f(\xi/|\xi|)\right>={\Gamma(m/2+p)\over\Gamma(m/2)}
\left<f(\xi/|\xi|)\right>.
\ees
This gives
\bes
<\Pi_iJ^\alpha\Pi_kJ^\beta>
={m\over 2}<\Pi_i\tilde J^\alpha\Pi_k \tilde J^\beta>,
\ees
\bes
<J^\alpha\Pi_kJ^\beta\Pi_i>
={m\over 2}<\tilde J^\alpha\Pi_k \tilde J^\beta\Pi_i>,
\ees
\bes
<|\xi|^2\Pi_iJ^\alpha\Pi_kJ^\beta>
={m(m+2)\over 4}<\Pi_i\tilde J^\alpha\Pi_k \tilde J^\beta>,
\ees
where
\bes
\tilde J^\alpha={J^\alpha\over |\xi|}
=a^{\alpha\beta}{\xi_\beta\over |\xi|}\,.
\ees
This allows to simplify the result somewhat:
\bea
\beta&=&
-{1\over 6m}\sum_{i=1}^s \mu^{-m/2}_i\tr_{\cal V}<\Pi_i>\bar a 
\nonumber\\[11pt]
&&
+{1\over 12(m-1)}
\sum_{1\le i,k\le s;\,i\ne k}
\left[\kappa_{ik}{(3m-2)}
+4(m+2)\mu_i\sigma_{ik}\right]
\tr_{\cal V}\left<\Pi_i\tilde J^\alpha \Pi_k\tilde J_\alpha\right>
\nonumber\\[11pt]
&&
-{1\over 2(m-1)}\sum_{1\le i,k\le s;\,i\ne k}
\kappa_{ik}\tr_{\cal V}
\left[\left<\tilde J^\nu\Pi_i \tilde J^\alpha \Pi_k\right>
+\left<\Pi_i \tilde J^\alpha \Pi_k \tilde J^\nu\right>\right]T_{[\nu\alpha]}.
\nonumber
\eea
Note that for {\it irreducible} representations the endomorphisms
$<\Pi_i>$, $\bar a=a^\mu{}_\mu\,$, and $<\Pi_iJ^\alpha\Pi_kJ_\alpha>$
are proportional to the identity $\II_{\cal V}$.

Finally, for the sake of completeness, we list the averages explicitly.
By using the formulas (\ref{HomGauss}) we get
\bea
<\Pi_i\tilde J_\alpha\Pi_k\tilde J_\beta>
&=&\sum_{1\le n,j\le s}{\Gamma(m/2)(2n+2j-2)!\over\Gamma(m/2+n+j-1)
2^{2n+2j-2}(n+j-1)!}\,
\nonumber\\[12pt]
&&\times 
g_{(\mu_1\mu_2}\cdots g_{\mu_{2n-3}\mu_{2n-2}}
g_{\nu_1\nu_2}\cdots g_{\nu_{2j-3}\nu_{2j-2}}g_{\gamma\delta)}\,
\nonumber\\[12pt]
&&\times
a^{(\mu_1\mu_2}\cdots a^{\mu_{2n-3}\mu_{2n-2}}a^\gamma{}_{\alpha}
a^{\nu_1\nu_2}\cdots a^{\nu_{2j-3}\nu_{2j-2}}a^{\delta)}{}_{\beta}\,.
\nonumber
\eea
Denoting by $\vee$ the symmetric product of symmetric forms
and by $\tr_g$ the total trace of a symmetric form, we can rewrite this
in the compact form
\bea
<\Pi_i\tilde J_\alpha\Pi_k\tilde J_\beta>
&=&\sum_{1\le n,j\le s}{\Gamma(m/2)(2n+2j-2)!\over\Gamma(m/2+n+j-1)
2^{2n+2j-2}(n+j-1)!}\,
\nonumber\\[12pt]
&&\times
\tr_g\,\left[(\vee^{n-1}a)\vee\tilde a_\alpha(\vee^{j-1}a)
\vee\tilde a_\beta\right]\,,
\nonumber
\eea
where $\tilde a_\alpha$ is a vector defined by 
$\tilde a_\alpha=a^{\mu}{}_\alpha\partial_\mu$.

This completes the general calculation of heat kernel
coefficient $\tr_{\cal V} a_1$. More explicit formulas may be obtained
in particular cases
by using explicit
formulas for the projections.

\section{The covariant semi-classical approximation}
\label{ten}
\setcounter{equation}0

In this section we show how the standard semi-classical method can be
adapted to compute the small-$t$ heat kernel asymptotics
of a non-Laplace type operator.
In treating the semi-classical approximation, 
we shall follow \cite{maslov76}.

The object of study is the fundamental solution of the heat equation
\bes
\left({1\over \varepsilon}\partial_t+F\right)U(t|x,x')=0
\ees
for a non-Laplace type operator 
\bes
F=-a^{\mu\nu}\nabla_\mu\nabla_\nu+q.
\ees
with a singular initial condition 
\be
U(t|x,x')\Big|_{t\to 0}=\delta(x,x').
\label{init}
\ee
Here $\varepsilon$ is a small formal parameter.

Let $\sigma^{\mu'}(x,x')$ denote the tangent vector to the geodesic 
connecting the points $x$ and $x'$ at the point $x'$, the norm of which is
equal to the length of that geodesic.
In this section we describe a systematic method for constructing the 
{\it local formal asymptotic solution} 
of the heat equation as $\varepsilon\to 0$. 
The initial condition suggests the following
Ansatz:
\bes
U(t|x,x')=J(t|x,x')\exp\left(-{1\over\varepsilon}S(t|x,x')\right)
\Omega(t|x,x'),
\ees
where $S$ is a scalar function, $J$ is another scalar function defined by
\bes
J(t|x,x')={1\over\sqrt{\det g_{\mu\nu}(x)}}\det\left[-
{1\over 2\pi\varepsilon}\nabla_\mu\nabla_{\nu'}S(t|x,x')\right]
{1\over\sqrt{\det g_{\mu\nu}(x')}},
\ees
and $\Omega(t)$ has the expansion 
\bes
\Omega(t|x,x')\sim\sum_{k\ge 0}\varepsilon^{k}\phi_k(t|x,x')
\ees
in powers of $\varepsilon$.
The leading asymptotics as $\varepsilon\to 0$ require
\bes
\left(\dot S +a^{\mu\nu}S_{;\mu}S_{;\nu}\right)\phi_{0}=0
\ees
where $\dot S=\partial_t S$ and $S_{;\mu}\equiv \nabla_\mu S$.
{}From this 
it follows that the function $S$ is determined by the {\it Hamilton-Jacobi} 
equation
\bes
\det_{\cal V}\left[\partial_t S 
+a^{\mu\nu}S_{;\mu}S_{;\nu}\right]=0.
\ees
Recalling the eigenvalues of the leading symbol,
we have
\bes
\prod_{i=1}^s\left[\partial_t S
+\mu_i g^{\mu\nu}S_{;\mu} S_{;\nu}\right]^{d_i}=0.
\ees
This has $s$ different solutions, one for each eigenvalue,
determined by
\be
{1\over\mu_i}\partial_t S_i+ g^{\mu\nu}S_{i;\mu}S_{i;\nu}=0.
\label{10-190}
\ee
These solutions differ by scaling $t\to t_i\equiv \mu_i t$,
i.e. $S(t|x,x')=S_0(\mu_i t|x,x')$, 
where $S_0$ is determined by the equation
\bes
\partial_t S_0+g^{\mu\nu}S_{0;\mu}S_{0;\nu}=0.
\ees
This is the Hamilton-Jacobi equation for a particle
moving in a curved manifold. 
There is a Hamiltonian 
system that corresponds to each Hamilton-Jacobi equation 
(\ref{10-190}). These Hamiltonian systems 
describe the geodesics parametrized
by $t_i=t\mu_i$.
The solution of this equation is given by 
the action along the geodesics connecting 
the points $x$ and $x'$ and parameterized so that
$x(0)=x'$, $x(t)=x$.
Thus
\bes
S_0(t|x,x')={\sigma(x,x')\over 2t}, 
\qquad S_i(t|x,x')={\sigma(x,x')\over 2t\mu_i}\,,
\ees
where $\sigma(x,x')$ is half the square of the length
of the geodesic.
This means that
\bes
J_i(t|x,x')=(4\pi t\mu_i\varepsilon)^{-m}\Delta(x,x'),
\ees
where $\Delta$ is the Van Vleck-Morette determinant.
As an immediate consequence of the Liouville Theorem, this satisfies
the equation
\bes
{1\over \mu_i}\partial_tJ_i+2\nabla_\mu(S^{;\mu}_iJ_i)=0,
\ees
or
\bes
\left(D+\sigma^{\mu}{}_{\mu}\right)J^{1/2}_i=0
\ees
where
\bes
D=t\partial_t+\sigma^\mu\nabla_\mu.
\ees
Here and below we denote the derivatives of $\sigma$ simply by 
adding indices to it, i.e. $\sigma_\mu\equiv \nabla_\mu \sigma$,
$\sigma_{\mu\nu}\equiv\nabla_\nu\nabla_\mu\sigma$, etc.
Recall that
\be
D\Delta^{1/2}={1\over 2}(m-\sigma^{\mu}{}_{\mu})\Delta^{1/2}.
\label{10-194}
\ee

Thus the semiclassical approximation as $\varepsilon\to 0$ {\it
polarizes} along different eigenvalues. That is, all quantities become
dependent on the eigenvalue, and the total solution is the
superposition of all particular solutions for all eigenvalues.
Thus, our final Ansatz is:
\bes
U(t|x,x')=\sum_{i=1}^s(4\pi t\mu_i\varepsilon)^{-m/2}
\Delta^{1/2}\exp\left(-{\sigma\over 2t\mu_i\varepsilon}\right)
\Omega_i(t|x,x'),
\ees
\bes
\Omega_i(t|x,x')\sim\sum_{k\ge 0}t^k\varepsilon^{k}\phi_{(i)k}(t|x,x').
\ees
For the function $\Omega_i(t)$, we get a transport equation
\bes
\left({1\over \varepsilon^2t^2}N_i
+{1\over \varepsilon t} L_i+M\right)\Omega_i(t)=0,
\ees
where
\bes
N_i={1\over 2\mu_i}\left(\sigma
-{1\over 2\mu_i}a^{\mu\nu}\sigma_\mu\sigma_\nu
\right),
\ees
\bes
L_i=t\partial_t-{m\over 2}
+{1\over \mu_i}\Delta^{-1/2}a^{\mu\nu}\sigma_\mu\nabla_\nu\Delta^{1/2}
+{1\over 2\mu_i}a^{\mu\nu}\sigma_{\mu\nu}\,,
\ees
\bes
M=\Delta^{-1/2}F\Delta^{1/2}\,.
\ees
The initial condition for $\Omega_i(t)$ is determined by the diagonal value of
the heat kernel
\be
\Omega_i(0|x,x)=<\Pi_i>,
\label{10-201}
\ee
where $<\Pi_i>$ is defined by (\ref{PiGauss}). While for irreducible 
representations this is proportional
to the identity matrix, for a general reducible bundle it is not.
By using the expansion of $\Omega_i(t)$,
we get recursion relations
\bea
N_i\phi_{(i)0}&=&0,\nonumber\\[10pt]
N_i\phi_{(i)1}&=&-L_i\phi_{(i)0}\,,\nonumber\\[10pt]
N_i\phi_{(i)k}&=&-(L_i+k-1)\phi_{(i)k-1}
-M\phi_{(i)k-2} \qquad k\ge 2.
\label{10-205}
\eea
Note that $N_i$ is just an endomorphism (i.e., has order zero
as a differential operator).

Now we have a quadratic form $a^{\mu\nu}\sigma_\mu\sigma_\nu$ which
can be expanded in terms of the same projections as before.
The role of the covector argument $\xi$ 
of each projection is now played by $\nabla\sigma$, and we denote
\bea
P_i\equiv\Pi_i(\nabla\sigma)&=&\sum_{n=0}^p \Pi_{i(2n)}^{\mu_1\cdots\mu_{2n}}
{\sigma_{\mu_1}\cdots\sigma_{\mu_{2n}}\over (2\sigma)^{n}}
\nonumber\\[12pt]
&=&\sum_{k=1}^s c_{ik} a^{(\mu_1\mu_2}\cdots a^{\mu_{2k-3}\mu_{2k-2})}
{\sigma_{\mu_1}\cdots \sigma_{\mu_{2k-2}}
\over (2\sigma)^{k-1}}
\,,
\eea
\bes
a^{\mu\nu}\sigma_\mu\sigma_\nu=2\sigma\sum_{i=0}^s\mu_i P_i\,.
\ees
Recall that $\nabla_\lambda a^{\mu\nu}=0$; this implies that 
the tensors $\Pi_{i(n)}$ (not $P_i$)
are covariantly constant:
\bes
\nabla\Pi_{i(n)}=0.
\ees
Note that this does not imply that the $P_i$ are covariantly constant,
since $\nabla\sigma$ is not.
Now, by using the decomposition 
of the leading symbol, we observe important properties
\bes
P_kN_i=N_iP_k= {\sigma\over 2\mu_i^2}(\mu_i-\mu_k)P_k\,,
\ees
\bes
P_iN_i=N_iP_i=0, \qquad (\II-P_i)N_i=N_i(\II-P_i)=N_i\,,
\ees
which imply that
\be
N_i=\sum_{0\le k\ne i\le s}{\sigma\over 2\mu_i^2}(\mu_i-\mu_k)P_k.
\label{10-210}
\ee
Next, decompose $\phi_{(i)k}$ according to the projection $P_i\,$: 
\bes
\phi_{(i)k}=\psi_{(i)k}+\chi_{(i)k},
\ees
where
\bes
\psi_{(i)k}=P_i\phi_{(i)k},\qquad
\chi_{(i)k}=(\II-P_i)\phi_{(i)k}.
\ees
Then the recursion (\ref{10-205}) takes the form
\bea
N_i\chi_{(i)0}&=&0,\nonumber
\\[10pt]
N_i\chi_{(i)1}&=&-L_i(\psi_{(i)0}+\chi_{(i)0}),\nonumber\\[10pt]
N_i\chi_{(i)k}&=&-(L_i+k-1)(\psi_{(i)k-1}+\chi_{(i)k-1})
-M(\psi_{(i)k-2}+\chi_{(i)k-2})\,.
\nonumber\\
&&
\label{10-215}
\eea
Now, by multiplying  this recursion by $P_n$, $n\ne i$, and using 
(\ref{10-210}), we obtain
\bea
{\sigma\over 2\mu_i^2}(\mu_i-\mu_n)P_n\chi_{(i)0}
&=&0,\nonumber\\[11pt]
{\sigma\over 2\mu_i^2}(\mu_i-\mu_n)P_n\chi_{(i)1}
&=&-P_nL_i\psi_{(i)0}\,,\nonumber\\[11pt]
{\sigma\over 2\mu_i^2}(\mu_i-\mu_n)P_n\chi_{(i)k}
&=&
-P_nL_i\psi_{(i)k-1}
-P_n(L_i+k-1)\chi_{(i)k-1}
\nonumber\\
&&-P_nM(\psi_{(i)k-2}+\chi_{(i)k-2})\,.
\nonumber\\&&
\label{10-218}
\eea
This recursion determines $\chi_{(i)k}$ {\it algebraically} in terms of
$\psi_{(i)k-1}$, $\psi_{(i)k-2}$, $\chi_{(i)k-1}$ and $\chi_{(i)k-2}$.
In particular, we find that
\bea
\chi_{(i)0}&=&0,\nonumber\\
\chi_{(i)1}&=&-{2\over\sigma}
\sum_{1\le n\ne i\le s}{\mu_i^2\over \mu_i-\mu_n}P_nL_i\psi_{(i)0}.
\nonumber
\eea

The recursion does not determine the $\psi_{(i)k}$ however. 
These are determined by
another {\it differential} recursion that is obtained by multiplying 
(\ref{10-215}) by $P_i\,$,
\bea
P_iL_i\psi_{(i)0}&=&0,\nonumber\\[11pt]
(P_iL_i+k)\psi_{(i)k}&=&-P_iL_i\chi_{(i)k}
-P_iM(\psi_{(i)k-1}+\chi_{(i)k-1}), \qquad k\ge 1.\nonumber
\eea

Now let us compute the operator $P_iLP_i$ entering this recursion.
We have
\bea
P_iL_iP_i
&=&t\partial_t
+{1\over \mu_i}P_i\Delta^{-1/2}a^{\mu\nu}\sigma_\mu P_i\nabla_\nu\Delta^{1/2}
\nonumber\\[11pt]
&&
+{1\over 2\mu_i}\left(P_ia^{\mu\nu}\sigma_\mu P_{i;\nu}
+P_ia^{\mu\nu}\sigma_{\mu\nu}P_i\right)
-{1\over 2}mP_i
\eea
where $P_{i;\nu}=\nabla_\nu P_i$.
Further, noting that
\bes
P_i a^{\mu\nu}\sigma_\mu\sigma_\nu P_i=2\mu_i\sigma P_i\,,
\ees
we get
\bes
P_ia^{\mu\nu}\sigma_\nu P_i=\mu_i\sigma^\mu P_i\,.
\ees
Taking into account the equation for the Van Vleck-Morette determinant 
(\ref{10-194}),
we obtain
\bes
P_iL_iP_i=P_i D+K_i,
\ees
where
\bes
D=t\partial_t+\sigma^\mu\nabla_\mu
\ees
is a first-order differential operator, and
\bes
K_i={1\over 2\mu_i}P_i\left[
2a^{\mu\nu}\sigma_\mu P_{i;\nu}
+(a^{\mu\nu}-\mu_ig^{\mu\nu})\sigma_{\mu\nu}\right]P_i
\ees
is some endomorphism.
Note that $\sigma^\mu=t{dx^\mu/dt}$. Therefore, the operator $D$ is
expressed in terms of the operator of {\it total} differentiation along the
geodesics,
\bes
D=t{d\over dt}.
\ees
The operator $D$ commutes with any function that depends only
on ``angular'' coordinates $\sigma^\mu/\sqrt\sigma$; 
in particular, it commutes with the projections:
\bes
DP_i=P_iD.
\ees
Thus the recursion for the $\psi$'s takes the form
\bea
(D+K_i)\psi_{(i)0}&=&0,\nonumber\\[11pt]
(D+k+K_i)\psi_{(i)k}&=&-P_iL_i\chi_{(i)k}
-P_iM(\psi_{(i)k-1}+\chi_{(i)k-1}), \qquad k\ge 1.
\nonumber\\
&&
\label{10-232}
\eea
But these are exactly the transport equations along geodesics.
They may be integrated with the appropriate initial 
conditions determined by (\ref{10-201}).
In particular, the first coefficient $\psi_{(i)0}$ is 
\bes
\psi_{(i)0}=\exp\left(-\int_0^t {d\tau\over\tau} K_i\right)<\Pi_i>,
\ees
where the integration is along the geodesic connecting the points $x'$ and $x$,
parametrized so that $x=x(\tau)$ that $x(0)=x'$, $x(t)=x$. 
In the standard case, i.e.\ for a Laplace type operator,
we have $K_i=0$. Therefore, the first coefficient is just $\phi_{0}=\II$.

Thus, the recursion relations determine the asymptotic solution
completely. The algorithm is the following: at each step determine first the
$\chi_{(i)k}$ by (\ref{10-218}), and then $\psi_{(i)k}$ by (\ref{10-232}).

\section{Concluding remarks}

Let us summarize the results of this paper.  We have studied in detail a 
general class of non-Laplace type operators, i.e.  elliptic second-order
partial differential operators acting on sections of a tensor-spinor vector
bundle over a compact manifold without boundary.  The only essential
assumptions that have been made are:  {\it i}) the positivity of the leading
symbol, $a^{\mu\nu}\xi_\mu\xi_\nu>0$  (in the sense of endomorphisms) for
$\xi\ne 0$, and  {\it ii}) the covariant constancy of the tensor $a^{\mu\nu}$,
i.e.\ $\nabla a=0$.   We constructed the leading order resolvent and the heat
kernel and computed the first two coefficients of the heat kernel asymptotic
expansion explicitly.   In the last section we developed an alternative
approach for the {\it off-diagonal} heat kernel  asymptotics for by making use
of the general theory of the semi-classical approximation.  This enabled us to
construct a new ansatz for  the heat kernel, as well as to find a complete set
of recursion relations for the coefficients of the {\it off-diagonal} 
asymptotic expansion. This generalizes a well known ansatz for the heat kernel
of Laplace type  operators (e.g. see \cite{avra91b}).  In contrast to the
Laplace type case, the off-diagonal  heat kernel for the non-Laplace type case
 exhibits some essentially new features, notably {\it polarization} along the
different eigenvalues of the leading symbol.   As an explicit example of a
non-Laplace type operator we considered the most general second-order operator
acting on the bundle of symmetric two-tensors.  We computed the  eigenvalues
of the leading symbol, the multiplicities and the corresponding projections.  


\end{document}